\documentclass[11pt, a4paper]{article}

\usepackage{graphicx}
\usepackage{bm}
\usepackage{amsmath, amssymb,theorem,verbatim}
\usepackage[round]{natbib} 
\usepackage{setspace}
\usepackage{multirow} 
\usepackage{hyperref}
\usepackage{algpseudocode}
\usepackage{textcomp}

\long\def\comment#1{}

\newtheorem{theorem}{Theorem}[section]

\newtheorem{lemma}{Lemma}[section]

\begin{document}


\author{Piotr Fryzlewicz\thanks{Department of Statistics, London School of Economics, Houghton Street, London WC2A 2AE, UK. Email: \url{p.fryzlewicz@lse.ac.uk}.}}




\title{Likelihood ratio Haar variance stabilization and normalization for Poisson and other non-Gaussian noise removal}

\oddsidemargin=0.25in
\evensidemargin=0in
\textwidth=6in
\headheight=0pt
\headsep=0pt
\topmargin=0in
\textheight=9in

\maketitle

\begin{abstract}
We propose a new methodology for denoising, variance-stabilizing and normalizing signals
whose both mean and variance are parameterized by a single unknown varying parameter, such as
Poisson or scaled chi-squared. Key to our methodology is the observation that the signed and square-rooted
generalized log-likelihood ratio test for the equality of the local means is approximately
and asymptotically distributed as standard normal under the null. We use these
test statistics within the Haar wavelet transform at each scale and location,
referring to them as the {\em likelihood ratio Haar (LRH) coefficients} of the data.
In the denoising algorithm, the LRH coefficients are used as thresholding decision statistics,
which enables the use of thresholds
suitable for i.i.d. Gaussian noise, despite the standard Haar coefficients of the signal being heteroscedastic.
In the variance-stabilizing and normalizing algorithm, the LRH coefficients replace the standard Haar
coefficients in the Haar basis expansion.
To the best of our knowledge, the variance-stabilizing and normalizing properties of the generalized
likelihood ratio test have not been interpreted or exploited in this manner before. We prove the consistency
of our LRH smoother for Poisson counts with a near-parametric rate, and various numerical experiments demonstrate
the good practical performance of our methodology.

\vspace{5pt}


\end{abstract}

\section{Introduction}

Wavelets have become an established tool in data science and were recently used in tasks
as diverse as seizure detection and epilepsy diagnosis
\citep{faaa15},
reconstruction of compressively sensed magnetic resonance images
\citep{zwjd14},
identification of protein-protein binding sites
\citep{jlx16},
wind power forecast
\citep{caz15},
and
digital image watermarking
\citep{aca15}.
Their popularity and potential for useful applications in data analysis
did not escape the attention of Peter Hall\footnote{This article is being submitted for publication in a
special issue of {\em Statistica Sinica} in memory of Prof. Peter Hall.}, who wrote, amongst
others, on
threshold choice in wavelet curve estimation \citep{hp96, hp96a},
wavelet methods for functions with many discontinuities \citep{hmt96},
wavelets for regression with irregular design \citep{ht97}
and block-thresholded wavelet estimators \citep{hkp99}.

In brief, traditional wavelet transformations are orthonormal transformations
of the input data (be it one-dimensional signals or images) into coefficients that 
carry information about the local behaviour of the data at a range of dyadic 
scales and locations. The major benefits of wavelet transforms is that they tend 
to offer sparse representation of the input data, with a small number of wavelet
coefficients often being able to encode much of the energy of the input signal,
and that they are computable and invertible in linear time via recursive pyramid
algorithms \citep{m89, daub}. Excellent reviews of the use of wavelets in statistics
can be found, for example, in \cite{vid} and \cite{n08}.
Because of their sparsity and rapid computability properties,
wavelets continue to be a useful statistical tool also in the ``big data'' era.

One canonical task facilitated by wavelets is the removal of noise from signals, which
usually proceeds by taking a wavelet transform of the data, thresholding away the
(typically many) wavelet coefficients that are small in magnitude, preserving those few
that are large in magnitude, and taking the
inverse wavelet transform. Since the seminal paper by \cite{dj94} in which the general
idea was first proposed, several other methods for wavelet smoothing of one-dimensional 
signals have appeared, including via block thresholding \citep{c99} or using the empirical
Bayes approach \citep{js03}. The majority of works in this area, including those listed
above, make the i.i.d. Gaussian noise assumption. By contrast, the focus of this article
is the treatment of signals in which the variance of the noise is a function of the mean;
this includes Poisson- or scaled-chi-squared-distributed signals. (Throughout the paper,
we refer to a distribution as a `scaled chi-squared', or simply `chi-squared',
if it takes the form $\sigma^2 m^{-1} \chi^2_m$.)

The simplest example of a wavelet transform, and the focus of this article, is the
Haar transform, which can be described as a sequence of symmetric scaled
differences of consecutive local means of the data, computed at dyadic scales and locations
and naturally forming a binary tree consisting of `parents' and `children'. Its local difference mechanism means that it offers
sparse representations for (approximately) piecewise-constant signals.

The starting point for this work is the observation that testing whether or not each
Haar coefficient of a signal exceeds a certain threshold (in the denoising task
described above) can be interpreted as the {\em likelihood ratio test} for the equality
of the corresponding local means of the signal in the i.i.d. Gaussian noise
model. Our aim in this paper is to take this observation further and propose
similar multiscale likelihood ratio tests for other distributions, most notably
those in which the variance is a function of the mean, such as Poisson or scaled
chi-squared. The proposed multiscale likelihood ratio tests will reduce to the traditional 
thresholding of Haar wavelet coefficients for Gaussian data, but will take entirely 
different and new forms for other distributions. This will lead to a new, unified class
of algorithms useful for problems such as e.g. Poisson intensity estimation, Poisson
image denoising, spectral density estimation in time series, or time-varying volatility
estimation in finance. (Although the focus of this article is on one-dimensional signals,
extension of our methodology to images is as straightforward as the extension of 
the standard one-dimensional Haar wavelet transform to two dimensions.)

The new multiscale likelihood ratio tests will naturally induce a new construction, {\em likelihood ratio (Haar) wavelets}, which 
have the benefit of producing (equivalents of) Haar wavelet coefficients
that are asymptotically standard normal under the null hypothesis of the corresponding
local means being equal, even for inhomogeneous non-Gaussian signals, such as
those listed above. This will
(a) make it much easier to choose a single threshold parameter in smoothing
these kinds of data and (b) serve as a basis for new normalizing transformations
for these kinds of data, which bring their distribution close to Gaussianity.
This article demonstrates both these phenomena.
Speaking heuristically, the particular device that enables these results
is the Wilks' theorem, according to which the signed square-rooted likelihood ratio statistic will often be approximately distributed as standard normal, an observation
that is key to our results but that, to the best of our knowledge, has not been
explored in a variance-stabilization context before.

Wavelet-based Poisson noise removal, with our without the use of a 
variance-stabilizing and/or normalizing transform,
has a long history. For a Poisson variable $X$, the \cite{ansc} transform $2(X + 3/8)^{1/2}$
brings its distribution to approximate normality with variance one. \cite{don} proposes
to pre-process Poisson data via the Anscombe transform, and then use wavelet-based
smoothing techniques suitable for i.i.d. Gaussian noise. This and a number of other
wavelet-based techniques for denoising Poisson-contaminated signals are reviewed and
compared in \cite{bdfs02}. These include the translation-invariant multiscale Bayesian 
techniques by \cite{kol5} and \cite{tn97, tn99}, shown to outperform earlier techniques 
in \cite{kol2, kol3} and \cite{nb99}. \cite{wn03} propose the use of ``platelets''
in Poisson image denoising. The Haar-Fisz methodology \cite{fiszhaar}, drawing
inspiration from earlier work by \cite{fisz1} outside the wavelet context, proceeds
by decomposing the Poisson data via the standard Haar transform, then variance-stabilizing
the Haar coefficients by dividing them by the MLE of their own standard deviation, and
then using thresholds suitable for i.i.d. Gaussian noise with variance one. Closely related
ideas appear in \cite{lvbu10} and \cite{rr10}. \cite{j06} extends the Haar-Fisz idea to other wavelets.
As an alternative to Anscombe's transform, which is known not to work well for 
low Poisson intensities,
\cite{zfs08} introduce a more involved square-root-type variance-stabilizing transform for (filtered)
Poisson data.
\cite{hw12} propose Bayesian Haar-based shrinkage for Poisson signals based on the exact distribution 
of the difference of two Poisson variates (the Skellam distribution).

In multiplicative set-ups, such as signals distrubuted as $X_k = \sigma_k^2 m^{-1} \chi^2_m$, the logarithmic
transform stabilizes the variance exactly, but does not bring the distribution of the transformed
$X_k$ close to normality, especially not for small values of $m$ such as 1 or 2. In the context of spectral density estimation in time series, in which
the signal is approximately exponentially distributed, wavelet shrinkage for the
logged (and hence variance-stabilized) periodogram is studied, amongst others, in \cite{m94},
\cite{g96}, \cite{pv04} and \cite{fovs10}. An alternative route, via pre-estimation of the variance
of the wavelet coefficients (rather than via variance stabilization) is taken in \cite{n96}.
Haar-Fisz or wavelet-Fisz estimation for the
periodogram or other (approximate) chi-squared models is developed in \cite{fssr06},
\cite{fn06} and
\cite{fnvs06}.

In more general settings, wavelet estimation for exponential families with quadratic or
cubic variance functions is considered in \cite{as01}, \cite{abs01} and \cite{bcz10}.
The Haar-Fisz or wavelet-Fisz transformations for unknown distributions are studied in
\cite{f08}, \cite{fdn07}, \cite{mnfr06} and \cite{n14}. Variance-stabilizing transformations
are reviewed in the (unpublished) manuscript by \cite{f09}.

Our approach departs from the existing literature in that our variance-stabilization and normalization
device does not involve either the pre-estimation of the variance (as, effectively, in the Haar-Fisz
transform) or the application of a Box-Cox-type transform (as in the Anscombe variance stabilization
for Poisson data or the logarithmic transform in multiplicative models). By contrast, our 
approach uses the entire likelihood for the purpose of variance-stabilization and normalization. To this end,
we exploit the ``variance-stabilizing'' and ``normalizing'' properties of the generalized likelihood ratio test, which to our
knowledge have not been interpreted or employed in the literature in this way and with this purpose before. As a result,
and in contrast to much of the existing literature, the thresholding decision in our proposed smoothing methodology
is not based on the usual wavelet detail
coefficients, but on the newly-proposed likelihood ratio Haar coefficients. For completeness, we mention that
\cite{kn04} construct multiscale decompositions of the Poisson likelihood, which leads them to consider binomial
likelihood ratio tests for the purpose of thresholding; however, this is done in a 
context that does not use the signed and square-rooted generalized log-likelihood ratio tests or utilize their 
variance-stabilizing or normalizing properties.

The paper is organized as follows. Section \ref{sec:genm} motivates and introduces the concept
of likelihood ratio Haar coefficients and outlines our general methodology for smoothing and
variance stabilization/normalization. Section \ref{sec:spec} describes our method in two special
cases, those of the Poisson and the scaled chi-squared distribution. Section \ref{sec:thpois}
formulates and discusses a consistency result for the Poisson smoother. 
Section \ref{sec:comp} compares and contrasts the likelihood ratio Haar coefficients and the Fisz coefficients.
Section \ref{sec:pp} provides a numerical study illustrating the practical performance
of our smoothing and variance stabilization/normalization algorithms. Section \ref{sec:ad}
offers a further discussion, and the proof of our theoretical result is in the appendix.

\section{General methodology}
\label{sec:genm}

We first describe the general setting. Let $X_1, \ldots, X_n$ be a sequence of
independent random variables such that
\[
X_k \sim F(\theta_k),
\]
where $F(\theta)$ is a family of distributions parameterized by a single unknown scalar
parameter $\theta$ such that $\mathbb{E}(X_k) = \theta_k$. The reader is invited to think of our two running examples:
$X_k \sim \mbox{Pois}(\lambda_k)$, and $X_k \sim \sigma_k^2 m^{-1} \chi^2_m$
(throughout the paper, we refer to the latter example as `scaled chi-squared' or simply
`chi-squared').
Extensions to higher-dimensional parameters are possible using similar devices
to that described in this paper, although certain aspects of the asymptotic normality 
are then lost, so we do not pursue this extension in the current work.

We recall the traditional Haar transform and the fundamentals of signal smoothing via 
(Haar) wavelet thresholding. In the following, we assume that $n = 2^J$,
where $J$ is an integer. Extensions to non-dyadic $n$ are possible, see e.g. \cite{w94}.
Given the input data $\mathbf{X} = (X_1, \ldots, X_{n})$, we define $\mathbf{s}_0 = 
(s_{0,1}, \ldots, s_{0,n}) = \mathbf{X}$. The Haar transform recursively performs the following steps
\begin{eqnarray}
\label{eq:s}
s_{j,k} & = & 2^{-1/2} (s_{j-1,2k-1} + s_{j-1,2k}),\\
\label{eq:d}
d_{j,k} & = & 2^{-1/2} (s_{j-1,2k-1} - s_{j-1,2k}),
\end{eqnarray}
for $j = 1, \ldots, J$ and $k = 1, \ldots, 2^{J-j}$. The indices $j$ and $k$ are thought
of as ``scale'' and ``location'' parameters, respectively, and the coefficients $s_{j,k}$ and $d_{j,k}$ as the 
``smooth'' and ``detail'' coefficients (respectively) at scale $j$, location $k$. It is easy to express $s_{j,k}$ and
$d_{j,k}$ as explicit functions of $\mathbf{X}$:
\begin{eqnarray*}
s_{j,k} & = & 2^{-j/2} \sum_{i=(k-1)2^j+1}^{k2^j} X_i, \\
d_{j,k} & = & 2^{-j/2} \left( \sum_{i=(k-1)2^j+1}^{(k-1)2^j+2^{j-1}} X_i - \sum_{i=(k-1)2^j+2^{j-1}+1}^{k2^j} X_i     \right).
\end{eqnarray*}
Defining $\mathbf{d}_j = (d_{j,k})_{k=1}^{2^{J-j}}$, the Haar transform $H$ of $\mathbf{X}$
is then
\[
H(\mathbf{X}) = (\mathbf{d}_1, \ldots, \mathbf{d}_J, s_{J,1}).
\]
The ``pyramid'' algorithm in formulae (\ref{eq:s}) and (\ref{eq:d}) enables the computation of 
$H(\mathbf{X})$ in $O(n)$ operations.
$H(\mathbf{X})$ is an orthonormal transform of $\mathbf{X}$ and can easily be inverted by undoing steps
(\ref{eq:s}) and (\ref{eq:d}). If the mean signal $\Theta = (\theta_1, \ldots, \theta_n)$ is piecewise-constant,
then those coefficients $d_{j,k}$ that correspond to the locally constant segments
of $\Theta$, being local differences of $\mathbf{X}$, will be zero-centered. 
This justifies the following
procedure for estimating the mean vector $\Theta$: take the Haar transform of $\mathbf{X}$,
retain those coefficients $d_{j,k}$ for which $|d_{j,k}| > t$ for a certain threshold $t$ and
set the others to zero, then take the inverse Haar transform of the thus-``hard''-thresholded vector.

In the i.i.d. Gaussian noise model, in which $X_k = \theta_k + \varepsilon_k$, where $\varepsilon \sim N(0,\sigma^2)$
with $\sigma^2$ assumed known, the operation $|d_{j,k}| > t$ is exactly the likelihood ratio test
for the local constancy of $\Theta$ in the following sense.
\begin{enumerate}
\item
Assume that the parameters $(\theta_i)_{i=(k-1)2^j+1}^{(k-1)2^j+2^{j-1}}$ are all constant and equal to $\theta^{(1)}$. The indices $i$ are the same as those corresponding to the $X_i$'s with positive weights in $d_{j,k}$.
\item
Assume that the parameters $(\theta_i)_{i=(k-1)2^j+2^{j-1}+1}^{k2^j}$ are all constant and equal to $\theta^{(2)}$. The indices $i$ are the same as those corresponding to the $X_i$'s with negative weights in $d_{j,k}$.
\item
Test $H_0\,\,:\,\,\theta^{(1)} = \theta^{(2)}$ against $H_1\,\,:\,\,\theta^{(1)} \neq \theta^{(2)}$; the Gaussian likelihood ratio test reduces to $|d_{j,k}| > t$, where $t$ is naturally related to the desired significance level. Note that $H_0$ can alternatively be phrased as $E(d_{j,k}) = 0$, and $H_1$ -- as $E(d_{j,k}) \neq 0$.
\end{enumerate}
Because under each $H_0$, the variable $d_{j,k}$ is distributed as $N(0,\sigma^2)$ due to the orthonormality of
the Haar transform, the same $t$ can meaningfully be used across different scales and locations $(j,k)$.

In models other than Gaussian, the operation $|d_{j,k}| > t$ can typically no longer be interpreted as
the likelihood ratio test for the equality of $\theta^{(1)}$ and $\theta^{(2)}$. Moreover, the distribution
of $d_{j,k}$ will not generally be the same under each $H_0$ but will, in many models, vary with the local
(unknown) parameters $(\theta_i)_{i=(k-1)2^j+1}^{k2^j}$, which makes the selection of $t$ in 
the operation $|d_{j,k}| > t$ challenging. This is, for example, the case in our running examples,
$X_k \sim \mbox{Pois}(\lambda_k)$ and $X_k \sim \sigma_k^2 m^{-1} \chi^2_m$, both of which are such
that $\mbox{Var}(X_k)$ is a non-trivial function of $E(X_k)$, which translates into the dependence of $d_{j,k}$ on the local
means vector $(\theta_i)_{i=(k-1)2^j+1}^{k2^j}$, even under the null hypothesis $E(d_{j,k}) = 0$.

In the (non-Gaussian) model under consideration, this can be remedied by replacing the operation
$|d_{j,k}| > t$ with a likelihood ratio test for $H_0\,\,:\,\,\theta^{(1)} = \theta^{(2)}$ against $H_1\,\,:\,\,\theta^{(1)} \neq \theta^{(2)}$
suitable for the distribution at hand. More specifically, denoting by $L(\theta\,|\,X_{k_1}, \ldots, X_{k_2})$ the likelihood of the constant parameter $\theta$ given the
data $X_{k_1}, \ldots, X_{k_2}$ and by $\hat{\theta}^{(1)}, \hat{\theta}^{(2)}$ the MLEs of $\theta^{(1)}, \theta^{(2)}$, respectively, we design a new Haar-like transform, in which we
replace the ``test statistic'' $d_{j,k}$ by

{\footnotesize
\begin{equation}
\label{eq:g}
g_{j,k} = \mbox{sign}(\hat{\theta}^{(1)} - \hat{\theta}^{(2)}) \left\{ 2\log\frac{\sup_{\theta^{(1)}}L(\theta^{(1)}\,|\,X_{(k-1)2^j+1}, \ldots, X_{(k-1)2^j+2^{j-1}})
\sup_{\theta^{(2)}}L(\theta^{(2)}\,|\,X_{(k-1)2^j+2^{j-1}+1}, \ldots, X_{k2^j})}
{\sup_{\theta}L(\theta\,|\,X_{(k-1)2^j+1}, \ldots, X_{k2^j})}    \right\}^{1/2},
\end{equation}
}

the signed and square-rooted generalized log-likelihood ratio statistic for testing $H_0$ against $H_1$.
The rationale is that by Wilks' theorem, under $H_0$, this quantity will asymptotically be distributed as $N(0,1)$ for a class of
models that includes, amongst others, our two running examples (Poisson and scaled chi-squared). We 
refer to $g_{j,k}$ as the {\em likelihood ratio Haar coefficient}
of $\mathbf{X}$ at scale $j$ and location $k$.

By performing this replacement, we tailor the Haar transform to the distribution of the input vector, rather than
using the standard Haar transform, which as we argued earlier is the most suitable for Gaussian input data.

\subsection{General methodology for smoothing}
\label{sec:sm}

We now outline the general methodology for signal smoothing (denoising) involving likelihood ratio Haar wavelets. The problem is
to estimate $\Theta$ from $\mathbf{X}$. Let $\mathbb{I}$ be the indicator function. The basic smoothing algorithm proceeds as follows.

\begin{enumerate}
\item
With $\mathbf{X}$ on input, compute the coefficients $s_{j,k}$, $d_{j,k}$ and $g_{j,k}$ as defined by (\ref{eq:s}), (\ref{eq:d}) and
(\ref{eq:g}), respectively.
\item
Estimate each $\mu_{j,k} := E(d_{j,k})$ by
\begin{equation}
\label{eq:llhest}
\hat{\mu}_{j,k} = \left\{
\begin{array}{ll}
0 & j=1, \ldots, J_0,\\
d_{j,k} \mathbb{I}(|g_{j,k}| > t) & j=J_0+1, \ldots, J.
\end{array}
\right.
\end{equation}
\item
Defining $\hat{\mu}_j = (\hat{\mu}_{j,k})_{k=1}^{2^{J-j}}$, compute the inverse Haar transform of the vector
$(\hat{\mu}_1, \ldots, \hat{\mu}_J, s_{J,1})$ and use it as the estimate $\hat{\Theta}$ of $\Theta$.
\end{enumerate}

The reason for setting $\hat{\mu}_{j,k} = 0$ at the finest scales is that a certain strong asymptotic normality
does not hold at these scales; see the proof of Theorem \ref{th:pois}. This theorem also specifies the permitted
magnitude of $J_0$.

The operation in the second line of (\ref{eq:llhest}) is referred to as hard thresholding; soft thresholding, in which the surviving
coefficients are shrunk towards zero, is also possible. The threshold $t$ is a tuning parameter of the procedure
and we discuss its selection later. The above algorithm differs from the standard smoothing using Haar wavelets
and hard thresholding in its use of the thresholding statistic: the standard Haar smoothing would use the decision
$|d_{j,k}| > t$, whereas we use $|g_{j,k}| > t$, as motivated in the introductory part of this section.

\subsection{General methodology for variance stabilization and normalization}
\label{ssec:stab}

Due to the fact that $g_{j,k}$ will typically be distributed as close to $N(0,1)$ under each $H_0$ (that is, for the
majority of scales $j$ and locations $k$), replacing the coefficients $d_{j,k}$ with $g_{j,k}$ 
can be viewed as ``normalizing'' or ``Gaussianizing''
the input signal in the Haar wavelet domain. The standard inverse Haar transform will then yield a normalized version
of the input signal. We outline the basic algorithm below.

\begin{enumerate}
\item
With $\mathbf{X}$ on input, compute the coefficients $s_{j,k}$ and $g_{j,k}$ as defined by (\ref{eq:s}) and
(\ref{eq:g}), respectively.
\item
Defining $\mathbf{g}_j = (g_{j,k})_{k=1}^{2^{J-j}}$, compute the inverse Haar transform of the vector
$(\mathbf{g}_1, \ldots, \mathbf{g}_J, s_{J,1})$ and denote the resulting vector by $G(\mathbf{X})$.
\end{enumerate}

Throughout the paper, we will be referring to $G(\mathbf{X})$ as the {\em likelihood ratio Haar transform} of $X$.

Inverting the standard Haar transform proceeds by transforming each pair of coefficients $(s_{j,k}, d_{j,k})$ into
$(s_{j-1,2k-1}, s_{j-1, 2k})$, hierarchically for $j = J, \ldots, 1$ (note that $s_{0,k} = X_k$). Similarly, to
demonstrate that the likelihood Haar transform is invertible, we need to show that it is possible
to transform $(s_{j,k}, g_{j,k})$ into $(s_{j-1,2k-1}, s_{j-1, 2k})$. This will be shown for the Poisson
case in Section \ref{ssec:pois} and for the chi-squared case in Section \ref{ssec:chisq}.

An (invertible) variance-stabilization transformation such as $G(\mathbf{X})$ is useful as it enables the smoothing
of $X$ in a modular way: (i) apply $G(X)$, (ii) use any smoother suitable for i.i.d. standard normal noise, (iii)
take the inverse of $G(X)$ to obtain a smoothed version of $X$.

\section{Specific examples: Poisson and chi-squared}
\label{sec:spec}

\subsection{The Poisson distribution}
\label{ssec:pois}

For $X_i \sim \mbox{Pois}(\lambda)$, we have $P(X_i = k) = \exp(-\lambda) \frac{\lambda^k}{k!}$
for $k = 0, 1, \ldots$, and if $X_s, \ldots, X_e \sim \mbox{Pois}(\lambda)$, then the
MLE $\hat{\lambda}$ of $\lambda$ is $\bar{X}_s^e = \frac{1}{e-s+1} \sum_{i=s}^e X_i$.
This, after straightforward algebra, leads to

{\footnotesize
\begin{eqnarray}
\label{eq:gpois}
g_{j,k} & = & \mbox{sign}(\bar{X}_{(k-1)2^j+1}^{(k-1)2^j+2^{j-1}} - \bar{X}_{(k-1)2^j+2^{j-1}+1}^{k2^j}) 2^{j/2}\\
& \times &
\left\{ \log(\bar{X}_{(k-1)2^j+1}^{(k-1)2^j+2^{j-1}})\bar{X}_{(k-1)2^j+1}^{(k-1)2^j+2^{j-1}} +  \log(\bar{X}_{(k-1)2^j+2^{j-1}+1}^{k2^j})\bar{X}_{(k-1)2^j+2^{j-1}+1}^{k2^j} - 
2\,\log(\bar{X}_{(k-1)2^j+1}^{k2^j}) \bar{X}_{(k-1)2^j+1}^{k2^j}  \right\}^{1/2},\nonumber
\end{eqnarray}
}
using the convention $0 \log\,0 = 0$.

We now show the invertibility of the Poisson likelihood ratio Haar transform. As mentioned in Section \ref{ssec:stab},
this amounts to showing that
$(s_{j,k}, g_{j,k})$ can be transformed into $(s_{j-1,2k-1}, s_{j-1, 2k})$. Denoting for brevity
$u = \bar{X}_{(k-1)2^j+1}^{(k-1)2^j+2^{j-1}}$, $v = \bar{X}_{(k-1)2^j+2^{j-1}+1}^{k2^j}$ and ignoring
some multiplicative constants and the square-root operation in $g_{j,k}$, which are irrelevant for 
invertibility, this amounts to showing that $(u,v)$ can be uniquely determined from $(u+v)/2$ and 
$\mbox{sign}(u-v)\{ u\log\,u + v\,\log\,v - (u+v)\,\log ((u+v)/2) \}$. The term 
$\mbox{sign}(u-v)$ determines whether $u\le v$ or vice versa, so assume that $u\le v$ w.l.o.g.
Denoting by $a$ the known value of $u+v$, observe that the function $(a-v)\log(a-v) + v\log\,v$ is
strictly increasing for $v \in [a/2, a]$, which means that $v$ can be determined uniquely and therefore
that the Poisson likelihood ratio Haar transform is invertible.

\subsection{The chi-squared distribution}
\label{ssec:chisq}

For $X_i \sim \sigma_i^2 m^{-1} \chi^2_m = \Gamma(m/2, m/(2\sigma_i^2))$, if $X_s, \ldots, X_e \sim 
\Gamma(m/2, m/(2\sigma^2))$, then the MLE $\hat{\sigma}^2$ of $\sigma^2$ is $\bar{X}_s^e = \frac{1}{e-s+1} \sum_{i=s}^e X_i$.
This, after straightforward algebra, leads to
{\footnotesize
\begin{eqnarray}
\label{eq:gchisq}
g_{j,k} & = & \mbox{sign}(\bar{X}_{(k-1)2^j+1}^{(k-1)2^j+2^{j-1}} - \bar{X}_{(k-1)2^j+2^{j-1}+1}^{k2^j}) 2^{j/2}\\
& \times &
\left\{ m \left[ \log(\bar{X}_{(k-1)2^j+1}^{k2^j}) - \frac{1}{2} \log(\bar{X}_{(k-1)2^j+1}^{(k-1)2^j+2^{j-1}}) 
-\frac{1}{2} \log(\bar{X}_{(k-1)2^j+2^{j-1}+1}^{k2^j}) \right] \right\}^{1/2}.\nonumber
\end{eqnarray}
}

Up to the multiplicative factor $m^{1/2}$, the form of the transform in (\ref{eq:gchisq})
is the same for any $m$, which means that the likelihood ratio Haar coefficients $g_{j,k}$
(computed with an arbitrary $m$) also achieve variance stabilization if $m$ is unknown
(but possibly to a constant different from one).

We now show the invertibility of the chi-squared likelihood ratio Haar transform. As in Section \ref{ssec:pois},
we denote $u = \bar{X}_{(k-1)2^j+1}^{(k-1)2^j+2^{j-1}}$, $v = \bar{X}_{(k-1)2^j+2^{j-1}+1}^{k2^j}$ and ignore
some multiplicative constants and the square-root operation in $g_{j,k}$ which are irrelevant for 
invertibility. Assume that $u\le v$ w.l.o.g. Denoting by $a$ the known value of $u+v$, observe that the function $-\log(a-v) - \log\,v$ is
strictly increasing for $v \in [a/2, a)$, which means that $v$ can be determined uniquely and therefore
that the chi-squared likelihood ratio Haar transform is invertible.

In both the Poisson (formula (\ref{eq:gpois})) and the chi-squared cases, $g_{j,k}$ is a function
of the local means $\bar{X}_{(k-1)2^j+1}^{(k-1)2^j+2^{j-1}}$ and $\bar{X}_{(k-1)2^j+2^{j-1}+1}^{k2^j}$,
which is unsurprising as these are sufficient statistics for the corresponding population means
in both these distributions.

Finally, we note that since these local means can be computed in computational time $O(n)$ using
the pyramid algorithm in formulae (\ref{eq:s}) and (\ref{eq:d}), the likelihood ratio Haar coefficients
$g_{j,k}$ can be also computed in linear time.

\section{Theoretical analysis of the likelihood ratio Haar Poisson smoother}
\label{sec:thpois}

In this section, we provide a theoretical mean-square analysis of the performance of the 
signal smoothing algorithm involving likelihood ratio Haar wavelets, described in Section
\ref{sec:sm}. Although we focus on the Poisson distribution, the statement of the result
and the mechanics of the proof will be similar for certain other distributions, including
scaled chi-squared. The following result holds.

\begin{theorem}
\label{th:pois}
Let $\Lambda = (\lambda_1, \ldots, \lambda_n)$ be a positive piecewise-constant vector, i.e. let there exist up to $N$ indices $\eta_1, \ldots, \eta_N$
for which $\lambda_{\eta_i} \neq \lambda_{\eta_i-1}$. Let $n = 2^J$, where $J$ is a positive integer.
Assume $\Lambda$ is bounded from above and away from zero, and denote
$\bar{\Lambda} := \max_i \lambda_i$, $\underline{\Lambda} := \min_i \lambda_i$,
$\Lambda' = \bar{\Lambda} - \underline{\Lambda}$ and
$\bar{\lambda}_s^e = \frac{1}{e-s+1} \sum_{i=s}^e \lambda_i$.
Let $X_k \sim \mbox{Pois}(\lambda_k)$ for $k = 1, \ldots, n$.
Let $\hat{\Lambda}$ be obtained as in the algorithm of Section \ref{sec:sm}, using threshold $t$ and with a fixed $\beta \in (0, 1)$
such that $J_0 = \lfloor \log_2 n^\beta \rfloor$. Then, 
with $d_{j,k}$ and $\mu_{j,k}$ defined in the algorithm of Section \ref{sec:sm} and with
$\bar{X}_s^e = \frac{1}{e-s+1} \sum_{i=s}^e X_i$,
on set ${\mathcal A} \cap {\mathcal B}$,
where
\begin{eqnarray*}
{\mathcal A} & = &  \{ \forall\,\, j = J_0+1, \ldots, J,\,\, k = 1, \ldots, 2^{J-j}\quad    (\bar{\lambda}_{(k-1)2^j+1}^{k2^j})^{-1/2}|d_{j,k} - \mu_{j,k}| < t_1 \}, \\
{\mathcal B} & = &  \{ \forall\,\, j = J_0, \ldots, J,\,\, k = 1, \ldots, 2^{J-j}\quad    2^{j/2}(\bar{\lambda}_{(k-1)2^j+1}^{k2^j})^{-1/2}
|\bar{X}_{(k-1)2^j+1}^{k2^j} - \bar{\lambda}_{(k-1)2^j+1}^{k2^j}|    <  t_2 \},
\end{eqnarray*}
whose probability approaches 1 as $n \to \infty$ if $t_1 = C_1 \log^{1/2} n$ and $t_2 = C_2 \log^{1/2} n$ with $C_1 > \{2(1-\beta)\}^{1/2}$ and
$C_2 > \{2(1-\beta)\}^{1/2}$, if threshold $t$ is such that
\begin{equation}
\label{eq:condt21}
t \ge \frac{t_1}{(1 - t_2 2^{-\frac{J_0+1}{2}} \underline{\Lambda}^{-1/2})^{1/2}},
\end{equation}
we have
\begin{eqnarray*}
\lefteqn{n^{-1}\| \hat{\Lambda} - \Lambda  \|^2 \le }\\
&&
\frac{1}{2}n^{-1} N (\Lambda')^2 (n^\beta - 1) +
2n^{-1}N\bar{\Lambda}^{1/2} \left\{   (J-J_0)(t^2 + t_1^2) \bar{\Lambda}^{1/2} + t^2 t_2 (2 + 2^{1/2})
n^{-\beta/2} \right\} + n^{-1} t_1^2 \bar{\lambda}_1^n,
\end{eqnarray*}
where $\| \cdot \|$ is the $l_2$-norm of its argument.
\end{theorem}

Bearing in mind the magnitudes of $t_2$ and $J_0$, we can see that the term $t_2 2^{-\frac{J_0+1}{2}} \underline{\Lambda}^{-1/2}$
becomes arbitrarily close to zero for large $n$, and therefore, from formula (\ref{eq:condt21}), the threshold constant $t$ can become arbitrarily close to $t_1$.
In particular, it is perfectly safe to set $t$ to be the ``universal'' threshold suitable for iid $N(0,1)$ noise \citep{dj94}, that is
$t = \{2\,\log\,n\}^{1/2}$. It is in this precise sense
that our likelihood ratio Haar construction achieves variance stabilization and normalization: in order to denoise Poisson signals in which the variance of the noise
depends on the local mean, it is now possible to use the universal Gaussian threshold, as if the noise were Gaussian with variance one.
Note that in classical Haar wavelet thresholding, which uses the thresholding decision $|d_{j,k}| > \tilde{t}$ rather than our $|g_{j,k}| > t$, 
the threshold $\tilde{t}$ would have to depend on the level of the Poisson intensity $\Lambda$ over the support of $d_{j,k}$, which is 
unknown; our approach completely circumvents this.

If the number $N$ of change-points does not increase with the sample size $n$, then the dominant term in the mean-square error is 
of order $O(n^{\beta-1})$.
This suggests that $\beta$ should be set to be ``arbitrarily small'', in which case the MSE becomes arbitrarily
close to the parametric rate $O(n^{-1})$.

\section{Links between likelihood ratio Haar wavelets and the Haar-Fisz methodology}
\label{sec:comp}

This section compares the likelihood ratio Haar coefficients $g_{j,k}$, defined in the general, Poisson and chi-squared
cases in formulae (\ref{eq:g}), (\ref{eq:gpois}) and (\ref{eq:gchisq}), respectively, to the Fisz coefficients $f_{j,k}$
\citep{fiszhaar}, which the above
work defines as the Haar coefficients $d_{j,k}$ divided by the maximum likelihood estimates of their own
standard deviation under the null hypothesis $E(d_{j,k}) = 0$. We start with the
Poisson case and note that by \cite{fiszhaar}, $f_{j,k}$ is then expressed as
\[
f_{j,k} = 2^{j/2-1} \frac{\bar{X}_{(k-1)2^j+1}^{(k-1)2^j+2^{j-1}} - \bar{X}_{(k-1)2^j+2^{j-1}+1}^{k2^j}}{\sqrt{\bar{X}_{(k-1)2^j+1}^{k2^j}}}.
\]
We first note that $\mbox{sign}(g_{j,k}) = \mbox{sign}(f_{j,k})$ and that Lemma \ref{lem:grt}, used with
$f(u) = u\,\log\, u; f(0) = 0$ in the notation of that lemma, reduces to $|g_{j,k}| \ge |f_{j,k}|$.
Moreover, since the inequality in Lemma \ref{lem:grt} arises as a simple application of Jensen's inequality
to the convex function $f(\cdot)$, it is intuitively apparent that the less convexity in $f(\cdot)$, the closer
$g_{j,k}$ will be to $f_{j,k}$. Noting that $f''(u) = u^{-1}$ and therefore the degree of convexity in $f(u)$ 
decreases as $u$ increases, it can heuristically be observed that $g_{j,k}$ and
$f_{j,k}$ should be closer to each other for larger values of 
$\bar{X}_{(k-1)2^j+1}^{(k-1)2^j+2^{j-1}}$ and $\bar{X}_{(k-1)2^j+2^{j-1}+1}^{k2^j}$ (i.e. for high Poisson
intensities), and further apart otherwise.

\begin{figure}[p]
\centering
  \includegraphics[width=0.6\linewidth]{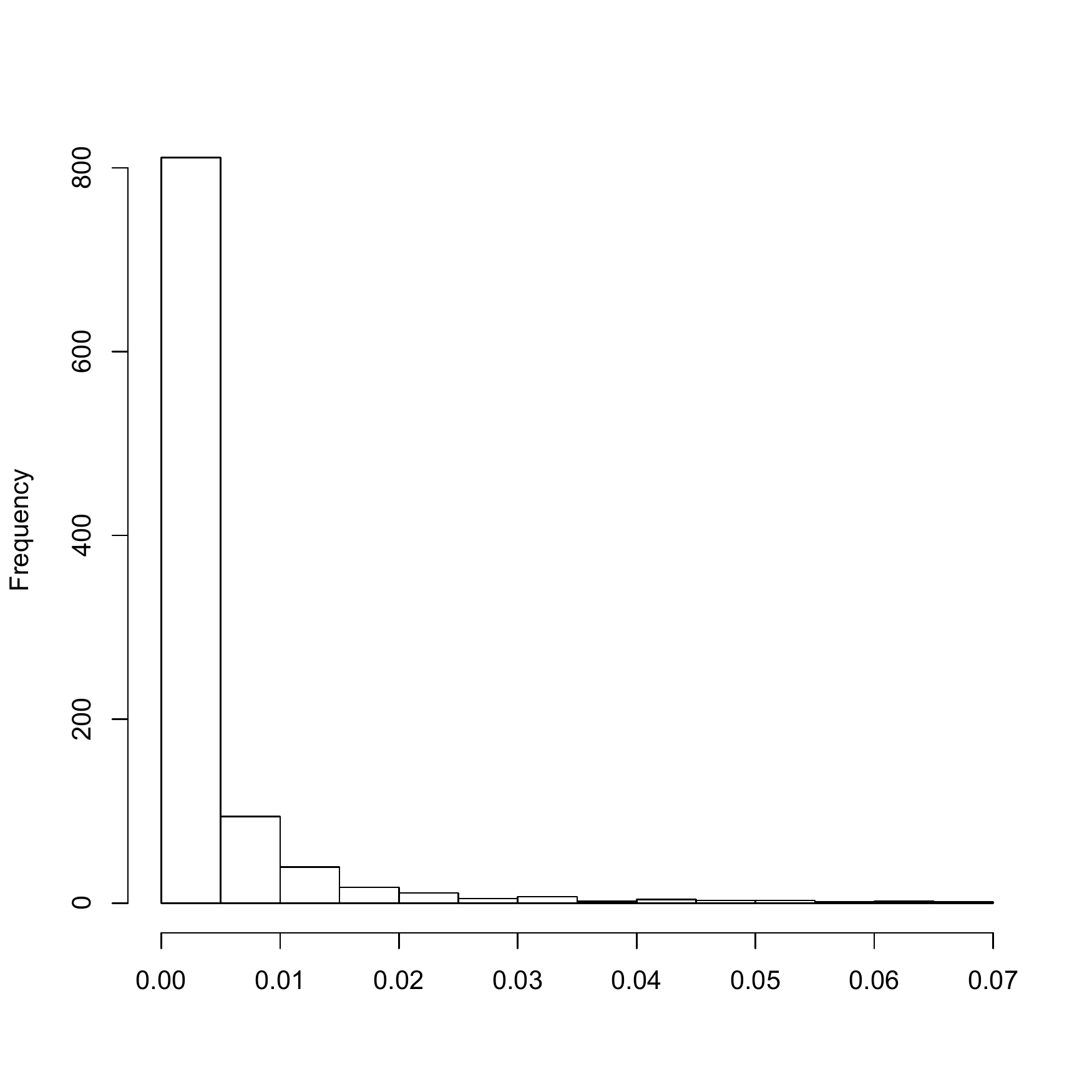}
  \caption{The Poisson case. Histogram of the empirical distribution of $\{|g_{j,k}^{(i)}| - |f_{j,k}^{(i)}|\}_{i=1}^{1000}$ with $j = 2$, $E(\bar{X}_{(k-1)2^j+1}^{(k-1)2^j+2^{j-1}}) = 10$, 
$E(\bar{X}_{(k-1)2^j+2^{j-1}+1}^{k2^j}) = 10.5$.}
  \label{fig:hist10}
\end{figure}

\begin{figure}[p]
\centering
  \includegraphics[width=0.6\linewidth]{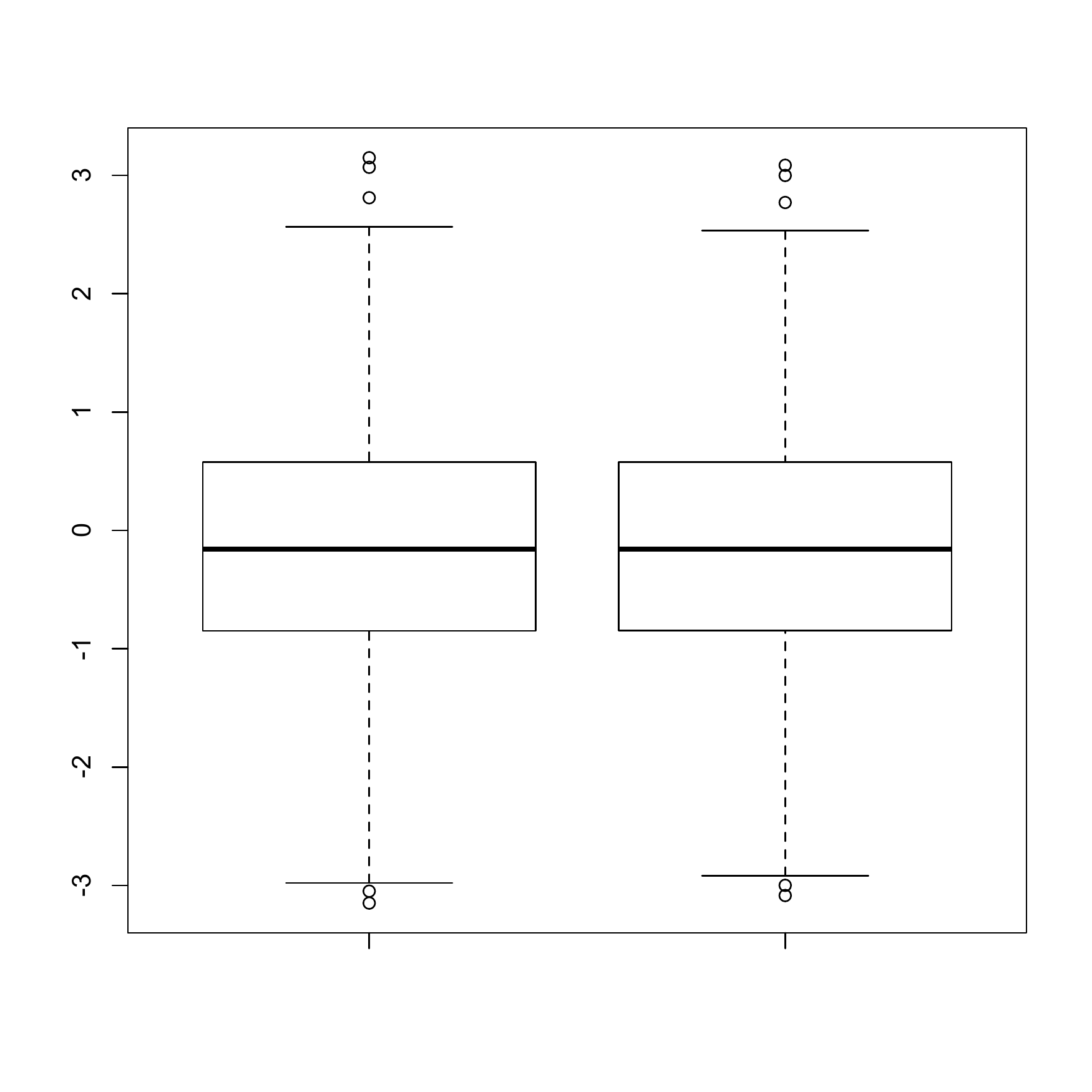}
  \caption{The Poisson case. Boxplots of the empirical distributions of $\{g_{j,k}^{(i)}\}_{i=1}^{1000}$ (left) and $\{f_{j,k}^{(i)}\}_{i=1}^{1000}$ (right) with $j = 2$, $E(\bar{X}_{(k-1)2^j+1}^{(k-1)2^j+2^{j-1}}) = 10$, 
$E(\bar{X}_{(k-1)2^j+2^{j-1}+1}^{k2^j}) = 10.5$.}
  \label{fig:box10}
\end{figure}

\begin{figure}[p]
\centering
  \includegraphics[width=0.6\linewidth]{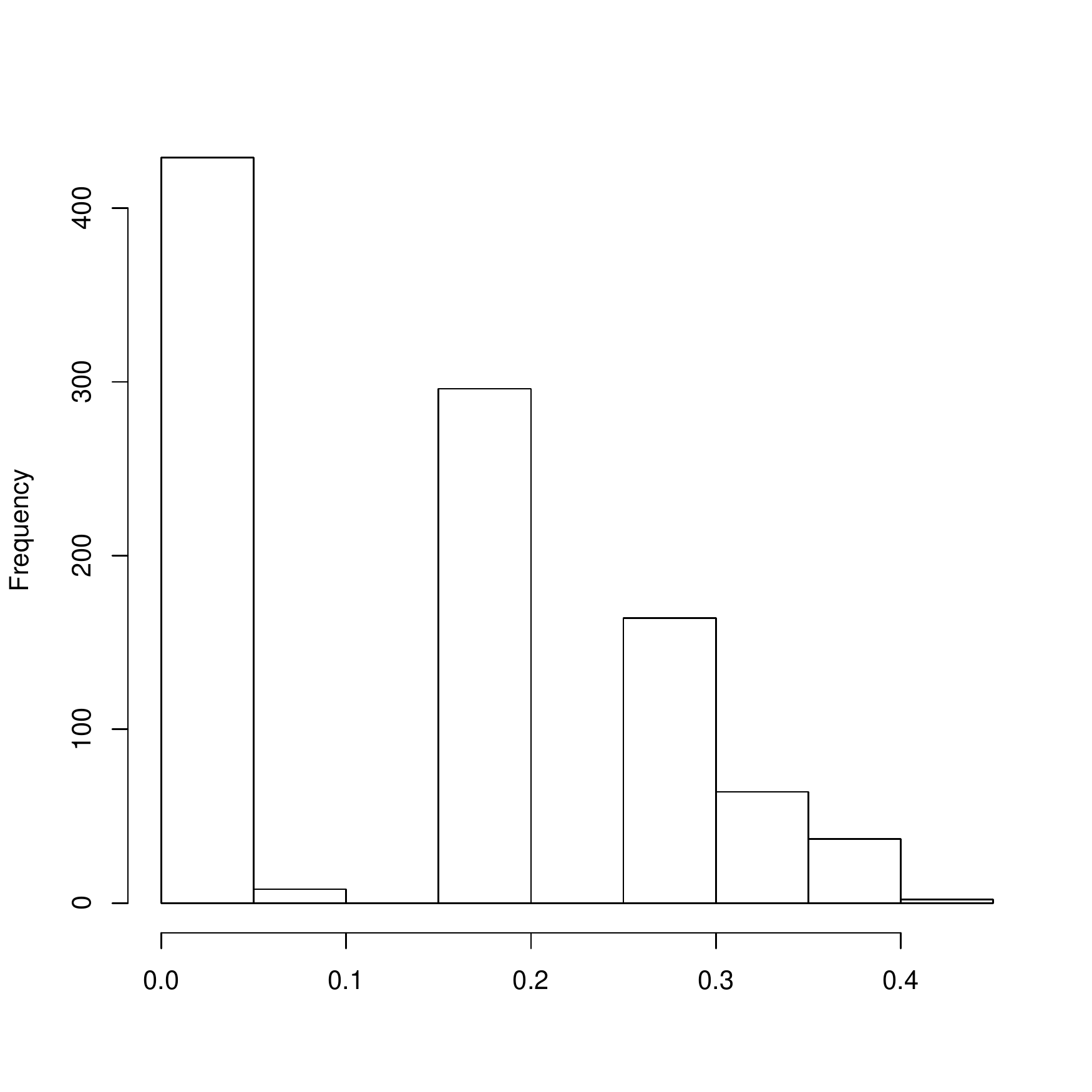}
  \caption{The Poisson case. Histogram of the empirical distribution of $\{|g_{j,k}^{(i)}| - |f_{j,k}^{(i)}|\}_{i=1}^{1000}$ with $j = 2$, $E(\bar{X}_{(k-1)2^j+1}^{(k-1)2^j+2^{j-1}}) = 0.2$, 
$E(\bar{X}_{(k-1)2^j+2^{j-1}+1}^{k2^j}) = 0.7$.}
  \label{fig:hist1}
\end{figure}

\begin{figure}[p]
\centering
  \includegraphics[width=0.6\linewidth]{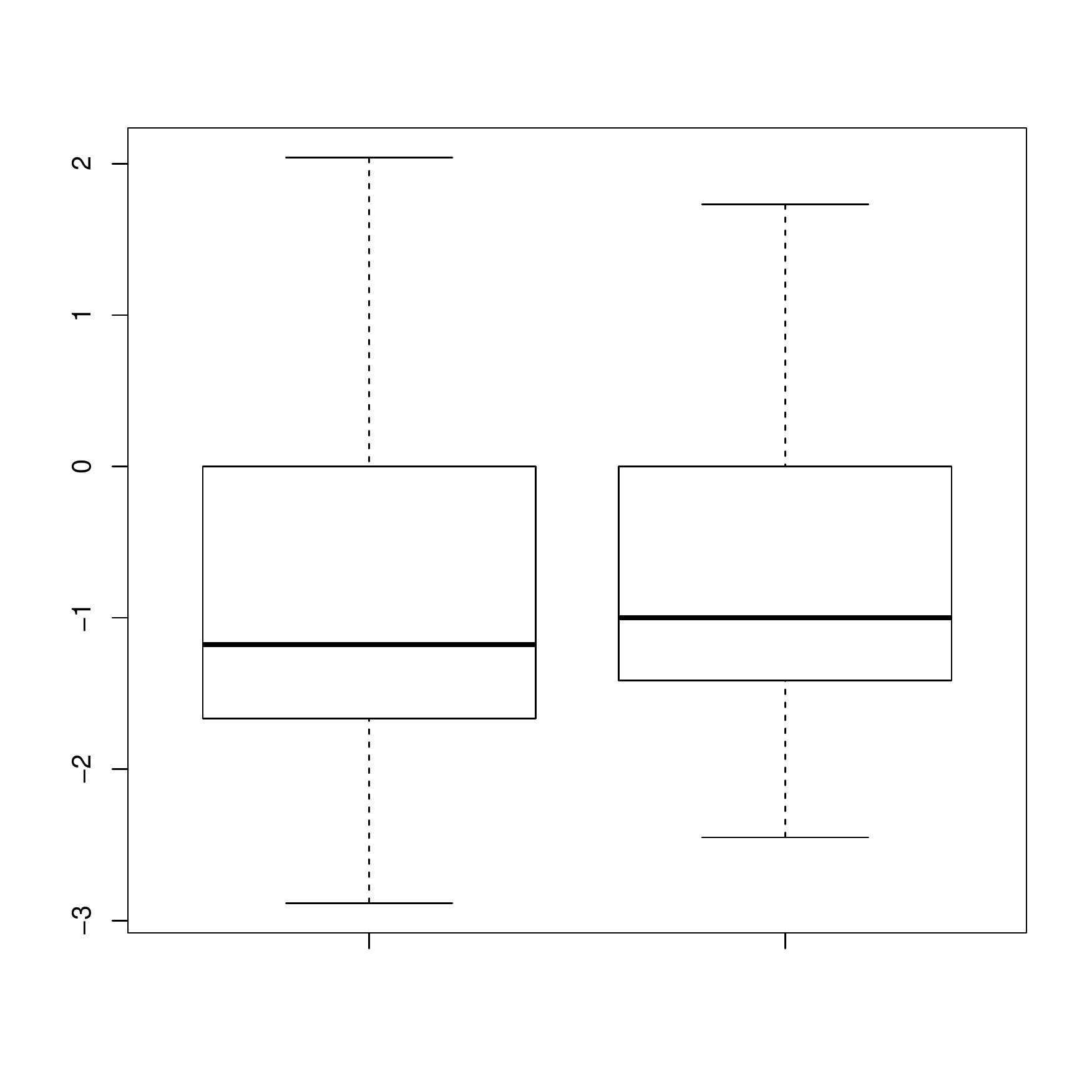}
  \caption{The Poisson case. Boxplots of the empirical distributions of $\{g_{j,k}^{(i)}\}_{i=1}^{1000}$ (left) and $\{f_{j,k}^{(i)}\}_{i=1}^{1000}$ (right) with $j = 2$, $E(\bar{X}_{(k-1)2^j+1}^{(k-1)2^j+2^{j-1}}) = 0.2$, 
$E(\bar{X}_{(k-1)2^j+2^{j-1}+1}^{k2^j}) = 0.7$.}
  \label{fig:box1}
\end{figure}

To illustrate this phenomenon and other interesting similarities and differences between the Fisz and
the likelihood ratio Haar coefficients in the Poisson case, consider the following two numerical examples, in which we simulate 1000 realisations of
$\bar{X}_{(k-1)2^j+1}^{(k-1)2^j+2^{j-1}}$ and $\bar{X}_{(k-1)2^j+2^{j-1}+1}^{k2^j}$ and compute the corresponding 1000 
realisations of $\{g_{j,k}^{(i)}\}_{i=1}^{1000}$ and $\{f_{j,k}^{(i)}\}_{i=1}^{1000}$.
\begin{itemize}
\item
$j = 2$, $E(\bar{X}_{(k-1)2^j+1}^{(k-1)2^j+2^{j-1}}) = 10$, 
$E(\bar{X}_{(k-1)2^j+2^{j-1}+1}^{k2^j}) = 10.5$. As is apparent from Figure \ref{fig:hist10},
the values of $g_{j,k}^{(i)} - f_{j,k}^{(i)}$ are close to zero. Figure \ref{fig:box10} provides
further evidence that the empirical distributions of $f_{j,k}^{(i)}$ and $g_{j,k}^{(i)}$ are difficult to distinguish by the
naked eye. Q-q plots (not shown) exhibit good agreement for both $g_{j,k}^{(i)}$ and $f_{j,k}^{(i)}$
with the normal distribution, and we have $\widehat{\mbox{Var}}(g_{j,k}^{(i)}) = 1.06$ and
$\widehat{\mbox{Var}}(f_{j,k}^{(i)}) = 1.05$, which provides evidence that both the likelihood ratio
Haar coefficients and the Fisz coefficients achieve good variance stabilization in this high-intensity case.
\item
$j = 2$, $E(\bar{X}_{(k-1)2^j+1}^{(k-1)2^j+2^{j-1}}) = 0.2$, 
$E(\bar{X}_{(k-1)2^j+2^{j-1}+1}^{k2^j}) = 0.7$. Figures \ref{fig:hist1} and \ref{fig:box1} demonstrate
that in this low-intensity case, the distributions of $f_{j,k}^{(i)}$ and $g_{j,k}^{(i)}$ are now further apart. The Fisz coefficients and
the likelihood ratio Haar coefficients seem to be similarly close to the normal distribution, with 
the empirical skewness and kurtosis for $f_{j,k}^{(i)}$ being 0.39 and 2.52 (respectively) and those for
$g_{j,k}^{(i)}$ being 0.35 and 2.53 (respectively). However, the likelihood ratio Haar coefficients achieve
far better variance stabilization in this low-intensity example: we have 
$\widehat{\mbox{Var}}(g_{j,k}^{(i)}) = 0.92$ versus $\widehat{\mbox{Var}}(f_{j,k}^{(i)}) = 0.68$.
\end{itemize}



We now turn to the chi-squared distribution. The Fisz coefficients for the $\sigma^2 \chi^2_1$ distribution are derived in \cite{fssr06},
those for
the exponential distribution ($\sigma^2 2^{-1} \chi^2_2$) appear in \cite{fnvs06} and the general case $\sigma^2 m^{-1} \chi^2_m$ is covered in 
\cite{f08}. In the general case of the $\sigma^2 m^{-1} \chi^2_m$ distribution, using the notation from the current paper, the Fisz coefficients
$f_{j,k}$ are expressed as
\begin{equation}
\label{eq:fchisq}
f_{j,k} = 2^\frac{j-3}{2} m^{1/2} \frac{\bar{X}_{(k-1)2^j+1}^{(k-1)2^j+2^{j-1}} - \bar{X}_{(k-1)2^j+2^{j-1}+1}^{k2^j}}{\bar{X}_{(k-1)2^j+1}^{k2^j}}.
\end{equation}

As in the Poisson case, we obviously have $\mbox{sign}(g_{j,k}) = \mbox{sign}(f_{j,k})$. Lemma \ref{lem:grt}, used with
$f(u) = -\log\, u$ in the notation of that lemma, reduces to $|g_{j,k}| \ge |f_{j,k}|$.
Moreover, by the same convexity argument as in the Poisson case, $g_{j,k}$ and
$f_{j,k}$ will be closer to each other for larger values of 
$\bar{X}_{(k-1)2^j+1}^{(k-1)2^j+2^{j-1}}$ and $\bar{X}_{(k-1)2^j+2^{j-1}+1}^{k2^j}$.

A major difference between the Poisson and the chi-square cases is that in the chi-square
case, $f_{j,k}$ is a compactly supported random variable (see formula (\ref{eq:fchisq})), whereas
$g_{j,k}$ is not. This difference does not apply in the Poisson case, in which neither $f_{j,k}$ nor $g_{j,k}$
are compactly supported. This has implications for how quickly $f_{j,k}$ and $g_{j,k}$ approach
the normal distribution (with increasing $j$ or $m$) in the chi-square case, and we illustrate this
numerically below.

\begin{figure}[p]
\centering
  \includegraphics[width=0.6\linewidth]{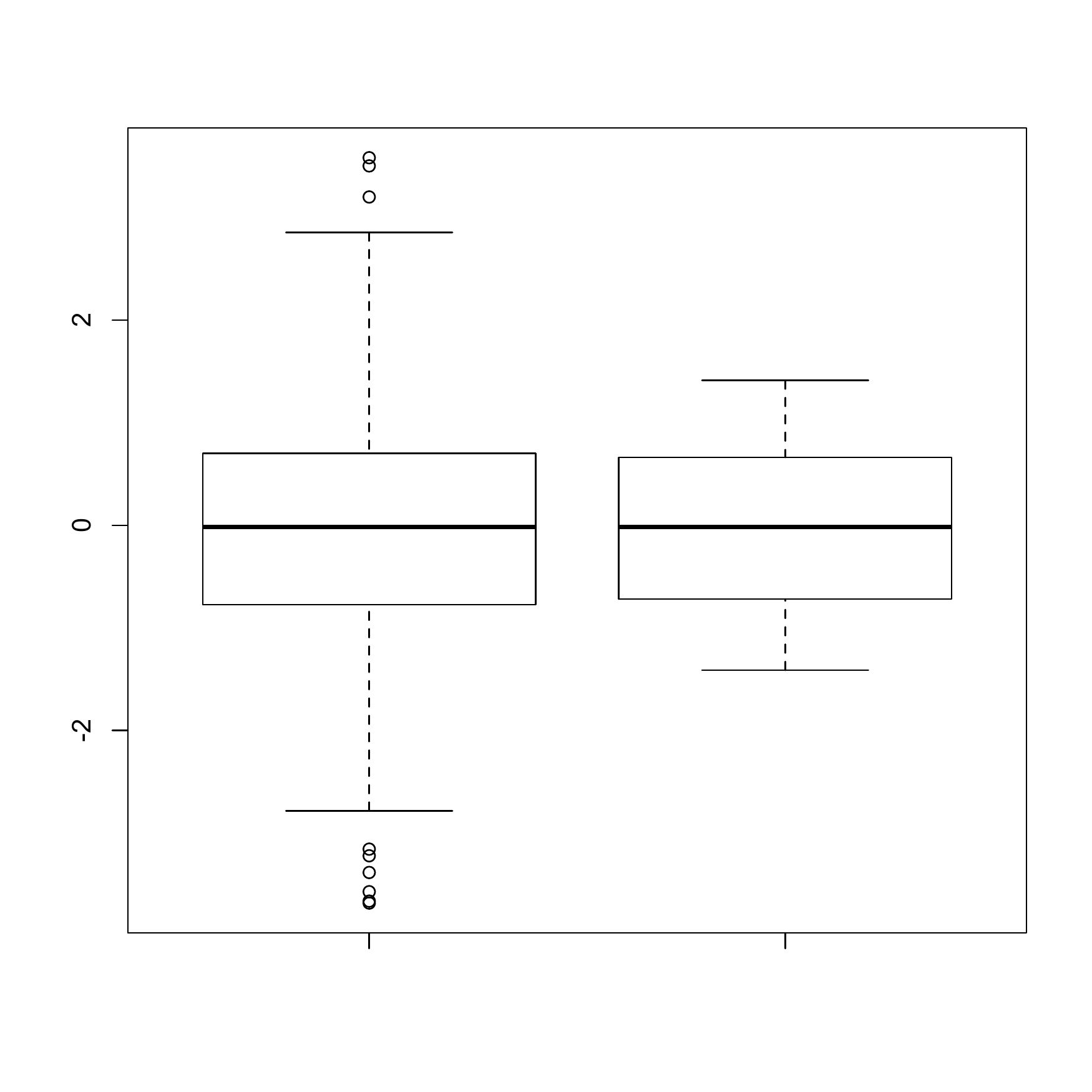}
  \caption{The chi-squared case. Boxplots of the empirical distributions of $\{g_{j,k}^{(i)}\}_{i=1}^{1000}$ (left) and $\{f_{j,k}^{(i)}\}_{i=1}^{1000}$ (right) with $m = 1$, $j = 2$, $E(\bar{X}_{(k-1)2^j+1}^{(k-1)2^j+2^{j-1}}) = 10$, $E(\bar{X}_{(k-1)2^j+2^{j-1}+1}^{k2^j}) = 10.5$.}
  \label{fig:box10_1}
\end{figure}

\begin{figure}[p]
\centering
  \includegraphics[width=0.6\linewidth]{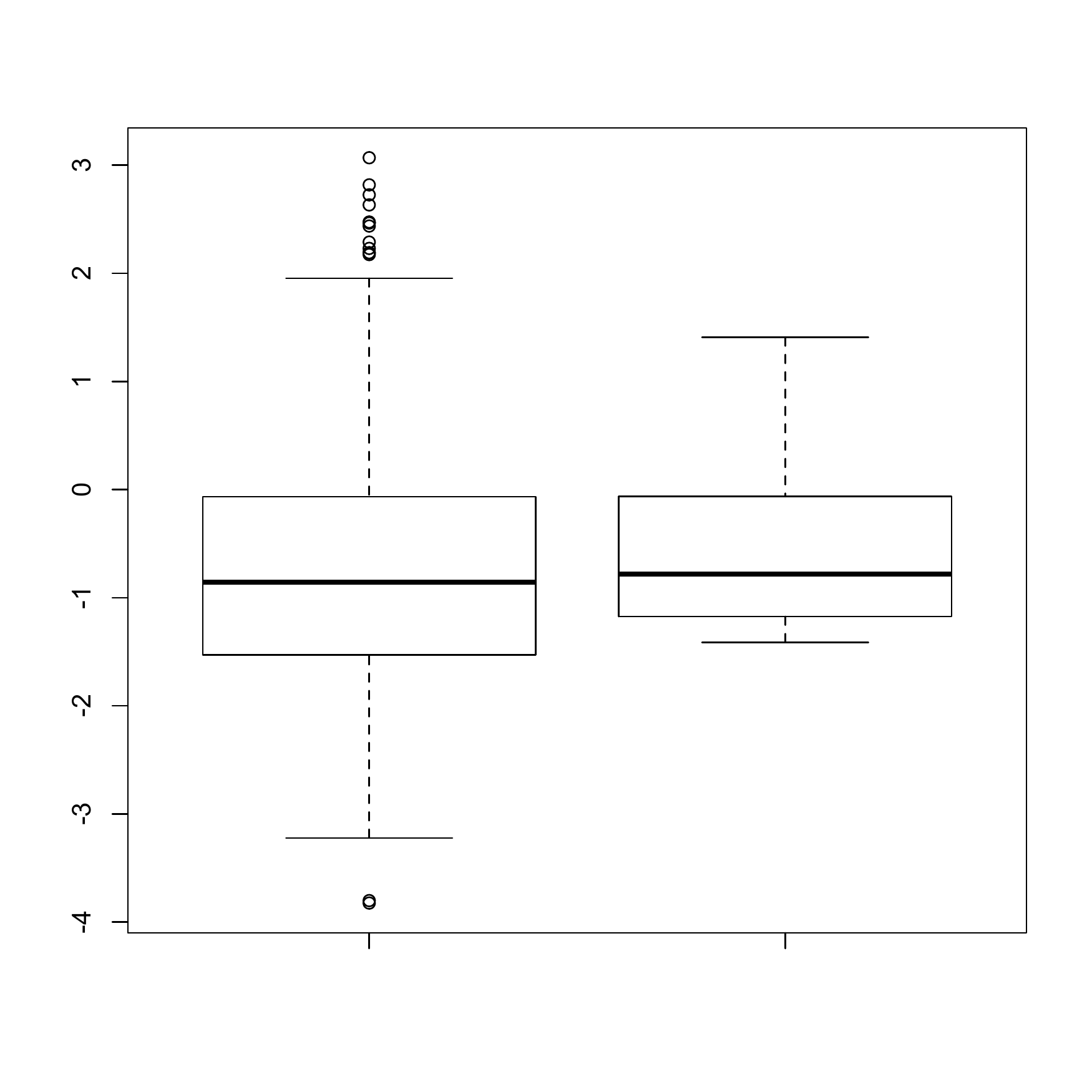}
  \caption{The chi-squared case. Boxplots of the empirical distributions of $\{g_{j,k}^{(i)}\}_{i=1}^{1000}$ (left) and $\{f_{j,k}^{(i)}\}_{i=1}^{1000}$ (right) with $m = 1$, $j = 2$, $E(\bar{X}_{(k-1)2^j+1}^{(k-1)2^j+2^{j-1}}) = 0.2$, $E(\bar{X}_{(k-1)2^j+2^{j-1}+1}^{k2^j}) = 0.7$.}
  \label{fig:box1_1}
\end{figure}

\begin{figure}[p]
\centering
  \includegraphics[width=0.6\linewidth]{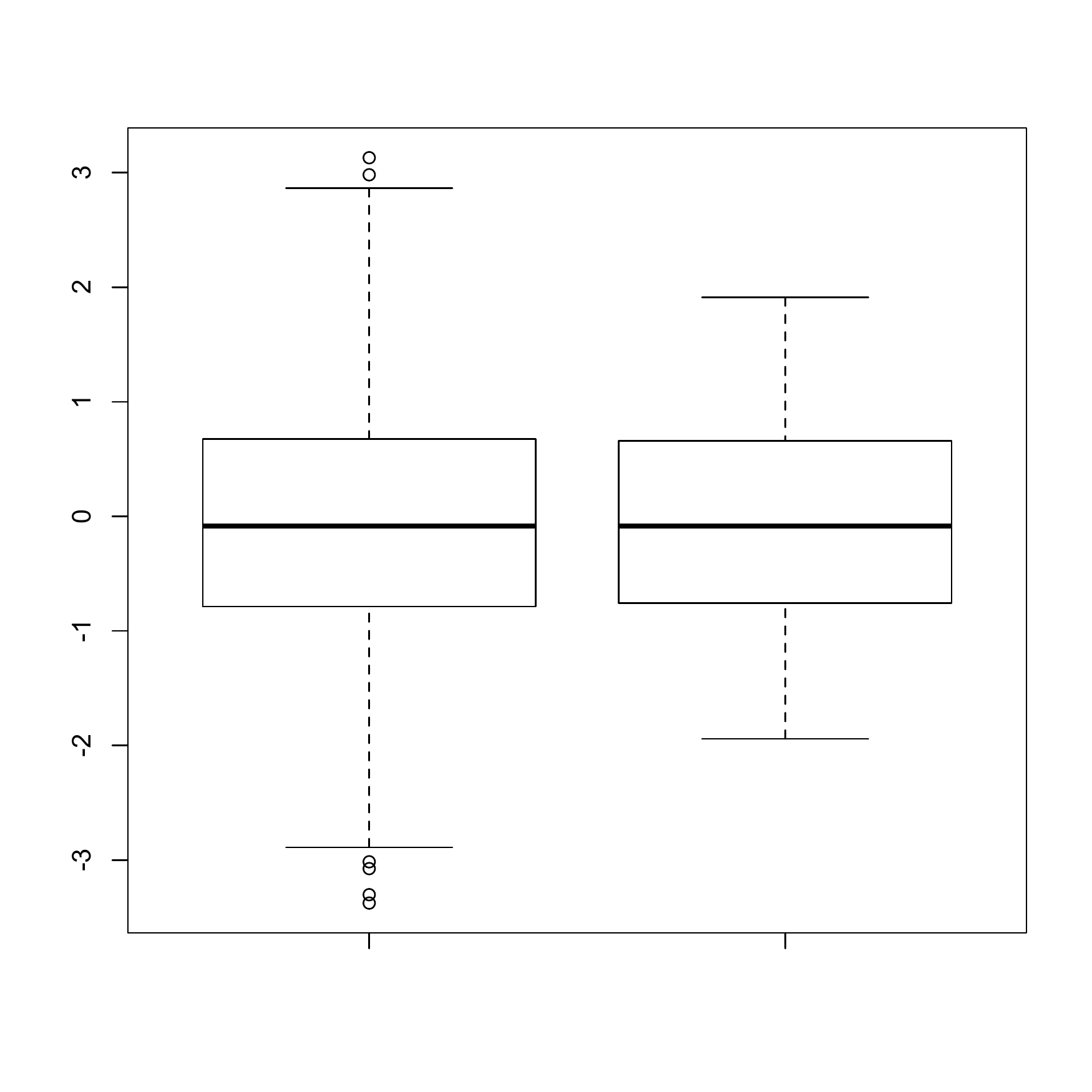}
  \caption{The chi-squared case. Boxplots of the empirical distributions of $\{g_{j,k}^{(i)}\}_{i=1}^{1000}$ (left) and $\{f_{j,k}^{(i)}\}_{i=1}^{1000}$ (right) with $m = 2$, $j = 2$, $E(\bar{X}_{(k-1)2^j+1}^{(k-1)2^j+2^{j-1}}) = 10$, $E(\bar{X}_{(k-1)2^j+2^{j-1}+1}^{k2^j}) = 10.5$.}
  \label{fig:box10_2}
\end{figure}

\begin{figure}[p]
\centering
  \includegraphics[width=0.6\linewidth]{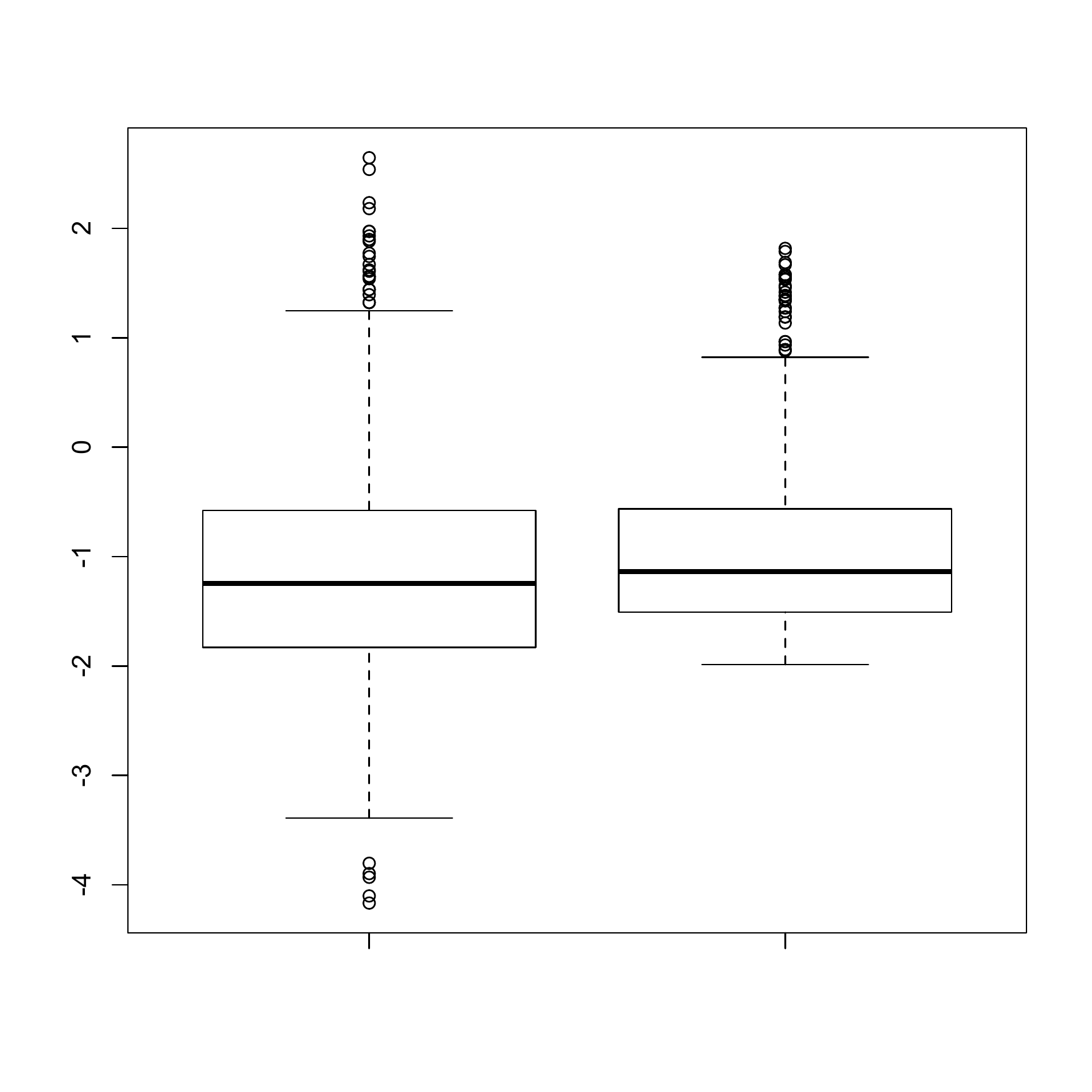}
  \caption{The chi-squared case. Boxplots of the empirical distributions of $\{g_{j,k}^{(i)}\}_{i=1}^{1000}$ (left) and $\{f_{j,k}^{(i)}\}_{i=1}^{1000}$ (right) with $m = 2$, $j = 2$, $E(\bar{X}_{(k-1)2^j+1}^{(k-1)2^j+2^{j-1}}) = 0.2$, $E(\bar{X}_{(k-1)2^j+2^{j-1}+1}^{k2^j}) = 0.7$.}
  \label{fig:box1_2}
\end{figure}

As before, we simulate 1000 realisations of
$\bar{X}_{(k-1)2^j+1}^{(k-1)2^j+2^{j-1}}$ and $\bar{X}_{(k-1)2^j+2^{j-1}+1}^{k2^j}$ and compute the corresponding 1000 
realisations of $\{g_{j,k}^{(i)}\}_{i=1}^{1000}$ and $\{f_{j,k}^{(i)}\}_{i=1}^{1000}$. We consider the following four cases.
\begin{itemize}
\item
$m = 1$, $j = 2$, $E(\bar{X}_{(k-1)2^j+1}^{(k-1)2^j+2^{j-1}}) = 10$, $E(\bar{X}_{(k-1)2^j+2^{j-1}+1}^{k2^j}) = 10.5$.
In this case, the likelihood ratio Haar coefficients provide far better variance stabilization and normalization than the
Fisz coefficients. For $f_{j,k}^{(i)}$, we have the following empirical values: variance 0.67, skewness 0.03, kurtosis 1.81.
For $g_{j,k}^{(i)}$, we have variance 1.29, skewness 0.03, kurtosis 3.06. Figure \ref{fig:box10_1} confirms the superiority
of the likelihood ratio Haar coefficients over the Fisz coefficients as regards their closeness to the normal distribution.
\item
$m = 1$, $j = 2$, $E(\bar{X}_{(k-1)2^j+1}^{(k-1)2^j+2^{j-1}}) = 0.2$, $E(\bar{X}_{(k-1)2^j+2^{j-1}+1}^{k2^j}) = 0.7$.
This low-sigma case differs from the previous one mainly in that both the likelihood ratio Haar coefficients and the
Fisz coefficients are skewed to the right, although the Fisz coefficients (much) more so.
For $f_{j,k}^{(i)}$, we have the following empirical values: variance 0.59, skewness 0.89, kurtosis 2.70.
For $g_{j,k}^{(i)}$, we have variance 1.23, skewness 0.46, kurtosis 3.1. Figure \ref{fig:box1_1} provides further visual
evidence of the higher degree of symmetry in the likelihood ratio Haar coefficients and its closeness to the normal
distribution.
\item
$m = 2$, $j = 2$, $E(\bar{X}_{(k-1)2^j+1}^{(k-1)2^j+2^{j-1}}) = 10$, $E(\bar{X}_{(k-1)2^j+2^{j-1}+1}^{k2^j}) = 10.5$.
As $m$ increases, both the likelihood ratio Haar coefficients and the Fisz coefficients move closer towards variance-one
normality, although again the likelihood ratio Haar coefficients beat Fisz.
For $f_{j,k}^{(i)}$, we have the following empirical values: variance 0.81, skewness 0.05, kurtosis 2.19.
For $g_{j,k}^{(i)}$, we have variance 1.16, skewness 0.03, kurtosis 2.97. Figure \ref{fig:box10_2} shows both empirical distributions.
\item
$m = 2$, $j = 2$, $E(\bar{X}_{(k-1)2^j+1}^{(k-1)2^j+2^{j-1}}) = 0.2$, $E(\bar{X}_{(k-1)2^j+2^{j-1}+1}^{k2^j}) = 0.7$.
In this low-sigma case also, the likelihood ratio Haar coefficients appear to be far closer to variance-one normality
than the Fisz coefficients.
For $f_{j,k}^{(i)}$, we have the following empirical values: variance 0.57, skewness 1.15, kurtosis 4.08.
For $g_{j,k}^{(i)}$, we have variance 1.04, skewness 0.45, kurtosis 3.64. Figure \ref{fig:box1_2} shows both empirical distributions.
\end{itemize}



Overall, our empirical observations from the above (and other unreported) numerical exercises are as follows.
For fine scales (i.e. those for which $j$ is small) and for low degrees of freedom $m$, the likelihood ratio Haar coefficients
are much closer to a normal variable with variance one than the corresponding Fisz coefficients. From the properties of the chi-squared
distribution, the effect of increasing $j$ while keeping $m$ constant is similar to the effect of increasing $m$ while keeping $j$ constant.
As $m$ or $j$ increases, the likelihood ratio Haar coefficients appear to move closer to the normal distribution with variance one. However,
for the same to happen with Fisz coefficients, the two means, 
$E(\bar{X}_{(k-1)2^j+1}^{(k-1)2^j+2^{j-1}})$ and $E(\bar{X}_{(k-1)2^j+2^{j-1}+1}^{k2^j})$, need to be relatively close to each other. The latter
phenomenon can also be observed in the Poisson case for increasing $j$. This is not unexpected as the results from \cite{fisz1}
suggest that the asymptotic normality with variance one arises when the two means approach each other
asymptotically; no results are provided in \cite{fisz1} on the case in which the two means diverge.

We end this section with an interesting interpretation of Lemmas \ref{lem:grt} and \ref{lem:mvt} in the case of the Poisson distribution. Note that
together, they imply
\[
2^{j/2-1} \frac{\left|\bar{X}_{(k-1)2^j+1}^{(k-1)2^j+2^{j-1}} - \bar{X}_{(k-1)2^j+2^{j-1}+1}^{k2^j}\right|}{\sqrt{\frac{2}{
\frac{1}{\bar{X}_{(k-1)2^j+1}^{(k-1)2^j+2^{j-1}}} + \frac{1}{\bar{X}_{(k-1)2^j+2^{j-1}+1}^{k2^j}}}
}} \ge
|g_{j,k}| \ge 2^{j/2-1} \frac{\left|\bar{X}_{(k-1)2^j+1}^{(k-1)2^j+2^{j-1}} - \bar{X}_{(k-1)2^j+2^{j-1}+1}^{k2^j}\right|}{\sqrt{\frac{1}{2}\left(
\bar{X}_{(k-1)2^j+1}^{(k-1)2^j+2^{j-1}} + \bar{X}_{(k-1)2^j+2^{j-1}+1}^{k2^j}\right)}},
\]
on in other words, the magnitude of the likelihood ratio Haar coefficient is bounded from below by the magnitude of the corresponding
Fisz coefficient and from above by the magnitude of a ``Fisz-like'' coefficient in which the arithmetic mean in the denominator has
been replaced by the harmonic mean.

\section{Practical performance}
\label{sec:pp}

\subsection{Likelihood ratio Haar smoothing}
\label{ssec:sm}

In Section \ref{sec:comp}, we demonstrated that the likelihood ratio Haar coefficients appeared to offer better normalization and variance stabilization
than the Fisz coefficients. In this section, we show that this translates into better MSE properties of the likelihood ratio Haar smoother than the analogous
Haar-Fisz smoother, in both the Poisson and the exponential models, on the examples considered. For comprehensive comparison of the performance
of the Haar-Fisz smoother to that of other techniques, see \cite{fiszhaar}, \cite{bdfs02} and \cite{f08}, among others.

Our test signals are [1] Donoho and Johnstone's (1994) \verb+blocks+ and [2] \verb+bumps+ functions, scaled to have (min, max) of
[1] $(0.681, 27.029)$ and [2] $(1, 12.565)$, both of length $n=2048$. We consider the following models:

\begin{description}
\item[(1a), (2a):] Poisson models, in which the signals [1], [2] (respectively) play the role of the Poisson intensity $\Lambda$, so that $X_k \sim \mbox{Pois}(\lambda_k)$.
\item[(1b), (2b):] Exponential models, in which the signals [1], [2] (respectively) play the role of the exponential mean $\sigma^2$, so that $X_k \sim \sigma^2_k\, \mbox{Exp}(1) = \sigma^2_k 2^{-1} \chi^2_2$.
\end{description}

\begin{figure}[p]
\centering
  \includegraphics[width=0.6\linewidth]{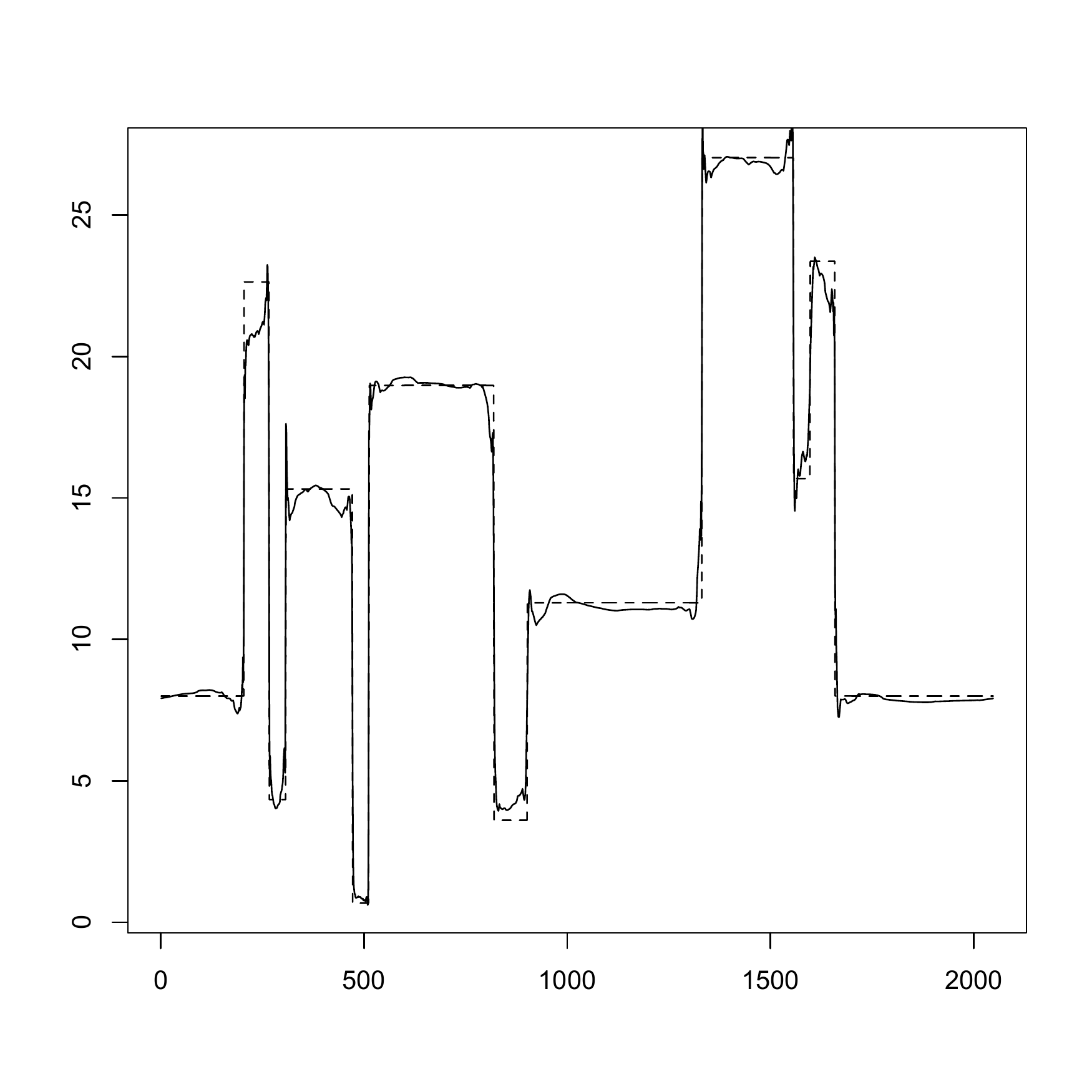}
  \caption{Sample likelihood ratio Haar reconstruction in model (1a), see Section \ref{ssec:sm} for details.}
  \label{fig:blp}
\end{figure}

\begin{figure}[p]
\centering
  \includegraphics[width=0.6\linewidth]{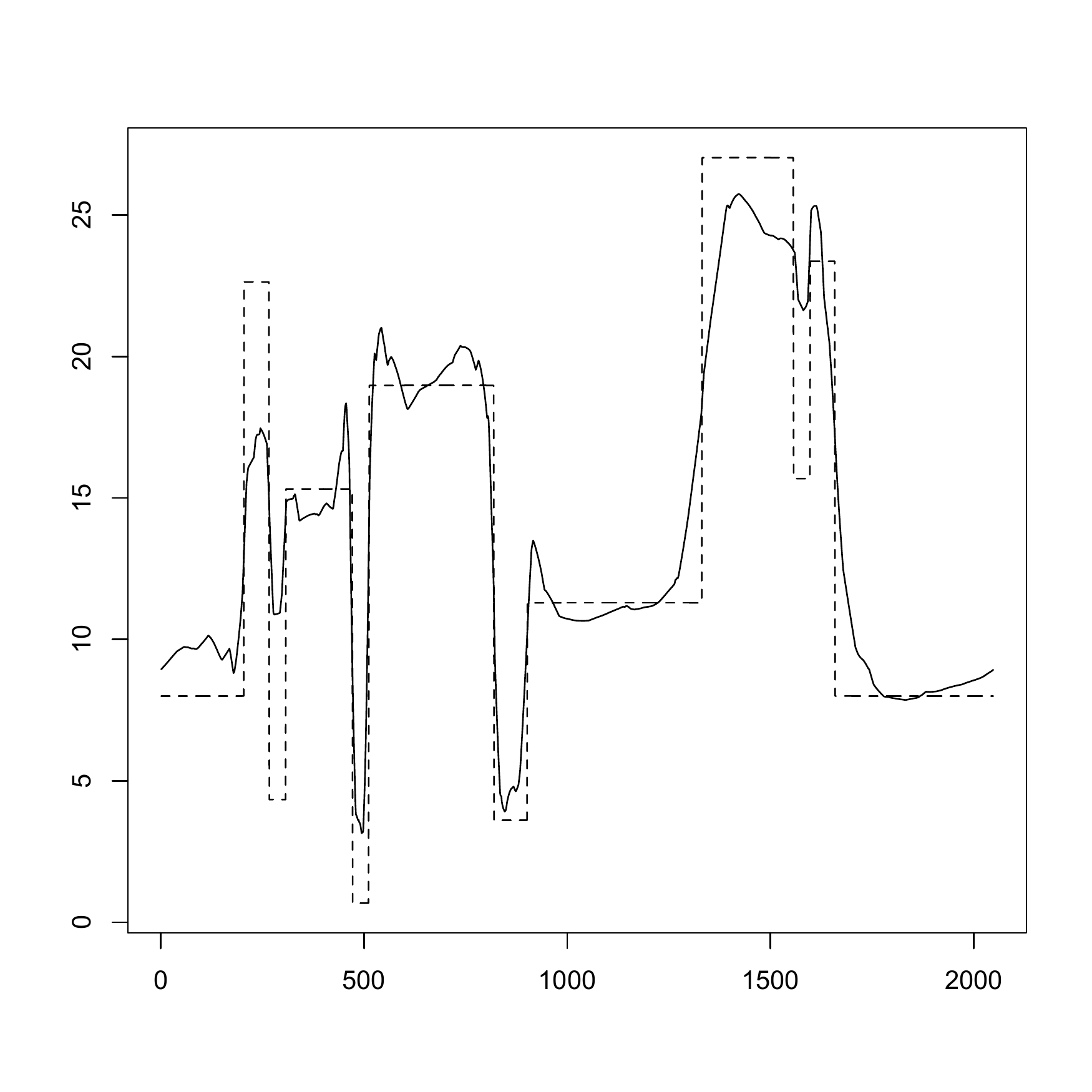}
  \caption{Sample likelihood ratio Haar reconstruction in model (1b), see Section \ref{ssec:sm} for details.}
  \label{fig:blc}
\end{figure}

\begin{figure}[p]
\centering
  \includegraphics[width=0.6\linewidth]{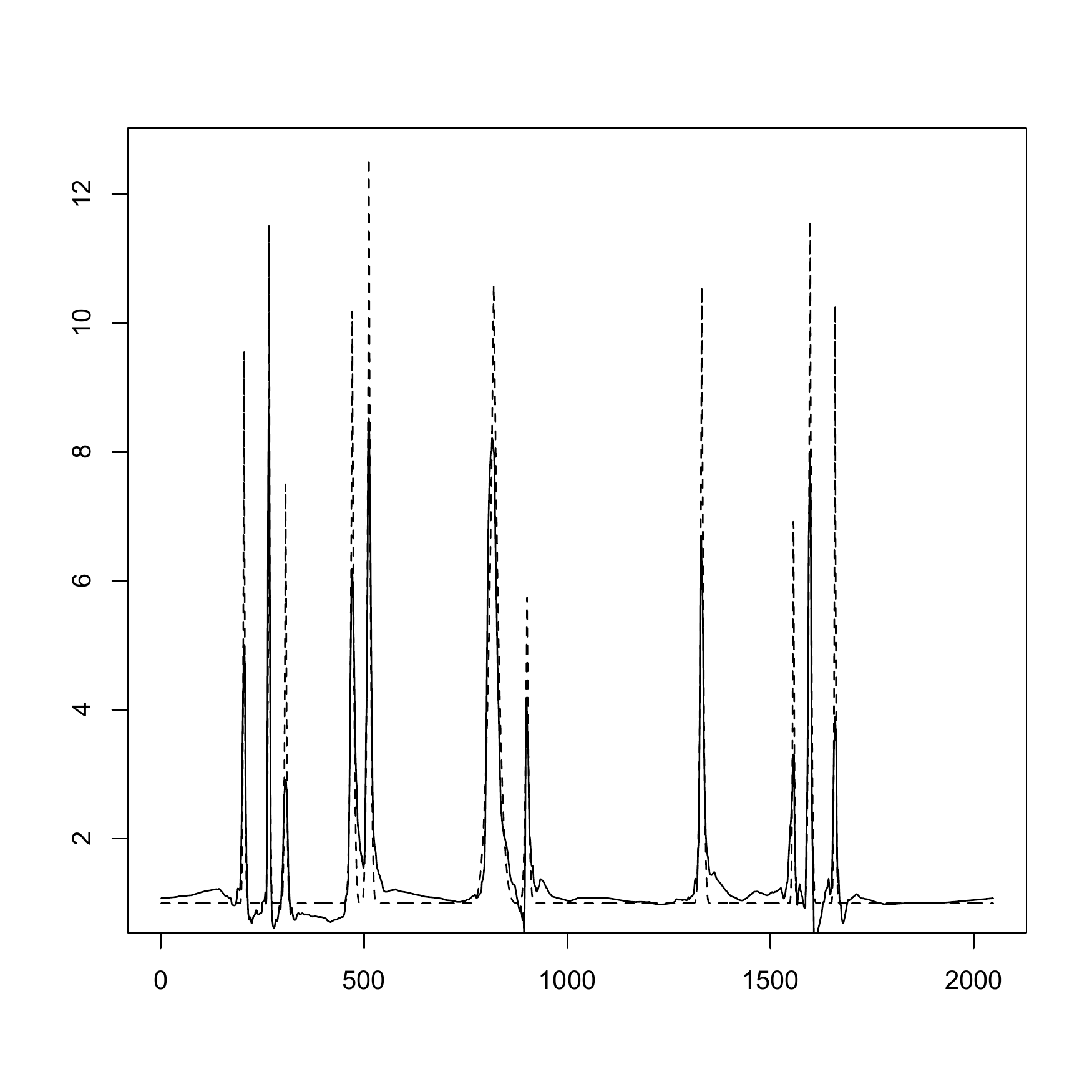}
  \caption{Sample likelihood ratio Haar reconstruction in model (2a), see Section \ref{ssec:sm} for details.}
  \label{fig:bup}
\end{figure}

\begin{figure}[p]
\centering
  \includegraphics[width=0.6\linewidth]{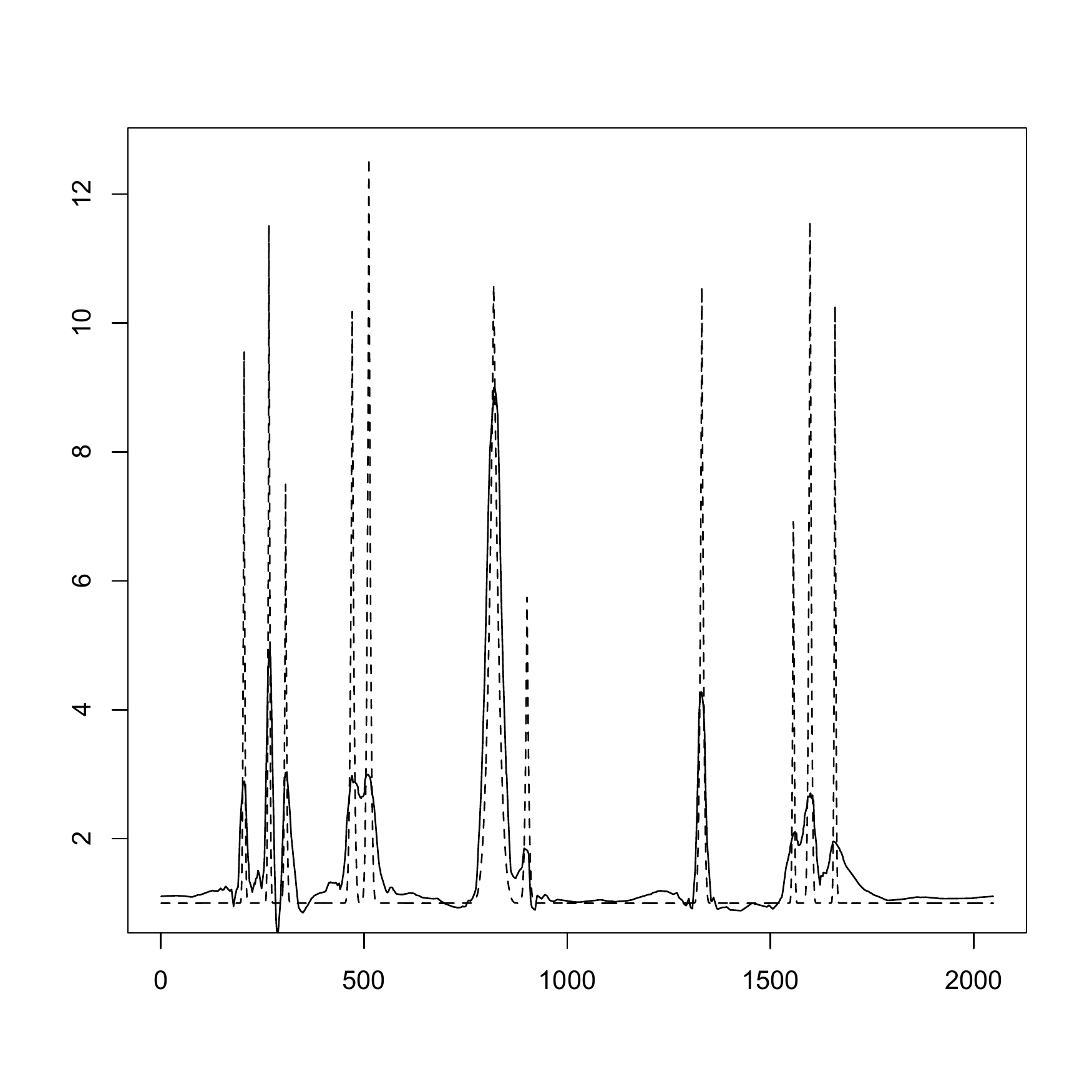}
  \caption{Sample likelihood ratio Haar reconstruction in model (2b), see Section \ref{ssec:sm} for details.}
  \label{fig:buc}
\end{figure}

For all models, we compare the MSE performance of ``like-for-like'' likelihood ratio Haar and Haar-Fisz smoothers, both constructed as described in 
Section \ref{sec:sm}, except the Haar-Fisz smoother uses the corresponding coefficients $f_{j,k}$ in place of $g_{j,k}$. We use the non-decimated
(translation invariant, stationary, maximum overlap) Haar transform \citep{nason2} to achieve fast averaging over all possibly cyclic shifts of the input data (note that the
classical decimated Haar transform as described in Section \ref{sec:genm} is not invariant to cyclic shifts). For better comparison of the effects of
thresholding, we use $J_0 = 0$ (i.e. we do not enforce any fine-scale coefficients to be zero other than via thresholding).
We use the universal threshold $t = \{2\,\log\,n\}^{1/2}$. Figures \ref{fig:blp}--\ref{fig:buc} shows sample reconstructions for the likelihood ratio Haar method.

Table \ref{tab:mise} shows that the likelihood ratio Haar smoother outperforms Haar-Fisz across all the models considered. For the Poisson models,
the improvement is fairly modest (2\% for \verb+blocks+, 4\% for \verb+bumps+) but for the exponential models, it is more significant 
(8\% for \verb+blocks+, 14\% for \verb+bumps+). One important reason for this improved performance is that as demonstrated earlier,
the likelihood ratio Haar coefficients have a higher magnitude than the corresponding Fisz coefficients, and therefore more easily survive
thresholding. This implies that the likelihood ratio Haar smoother lets through ``more signal'' compared to the Haar-Fisz smoother if both use
the same threshold, however chosen. Another possible reason is that as shown in Section \ref{sec:comp}, the likelihood ratio Haar coefficients
are closer to variance-one normality than the Fisz coefficients and therefore the use of thresholds designed for standard normal noise
may be more suitable for them.



\begin{table}
\centering
  \begin{tabular}{|c|c|c|c|c|}
    \hline
    Method \textbackslash\,Model & (1a) & (1b) & (2a) & (2b) \\
    \hline
Haar-Fisz    & 0.615 & 8.647 & 0.357 & 1.053 \\
Likelihood ratio Haar  & 0.605 & 7.958 & 0.341 &  0.905 \\
        \hline
  \end{tabular}
\caption{MSE over 1000 simulations for the two methods and four models described in Section \ref{ssec:sm}.\label{tab:mise}}
\end{table}

We end this section with two remarks:
\begin{enumerate}
\item
We use the simple universal threshold only to be able to demonstrate and explain the difference between both techniques in clear terms. It is likely
that their performance can be improved further if more elaborate thresholding techniques suitable for i.i.d. Gaussian data are applied 
instead. Their selection can be combined with a suitable choice of $J_0$ (rather than simply setting $J_0=0$, as is done here.)
\item
We use the non-decimated Haar transform for better MSE performance. This, however, uses ``circular'' boundary conditions, meaning that the value
of the signal at the left edge is assumed to equal that at the right edge. If accuracy at the right edge is important (e.g. for the purpose of forecasting or
extrapolating into the future), the standard decimated Haar transform may be used instead.
\end{enumerate}

\subsection{Likelihood ratio Haar transform}

In this section, we briefly illustrate the normalizing and variance-stabilizing properties of the likelihood ratio Haar transform
$G(\cdot)$ described in Section \ref{ssec:stab}, using data simulated from models (1a) and (1b) of Section \ref{ssec:sm} (i.e. the 
\verb+blocks+ signal with Poisson and exponential noise, respectively). As in that section, we use the non-decimated version of the
Haar transform.

\begin{figure}[p]
\centering
  \includegraphics[width=\linewidth]{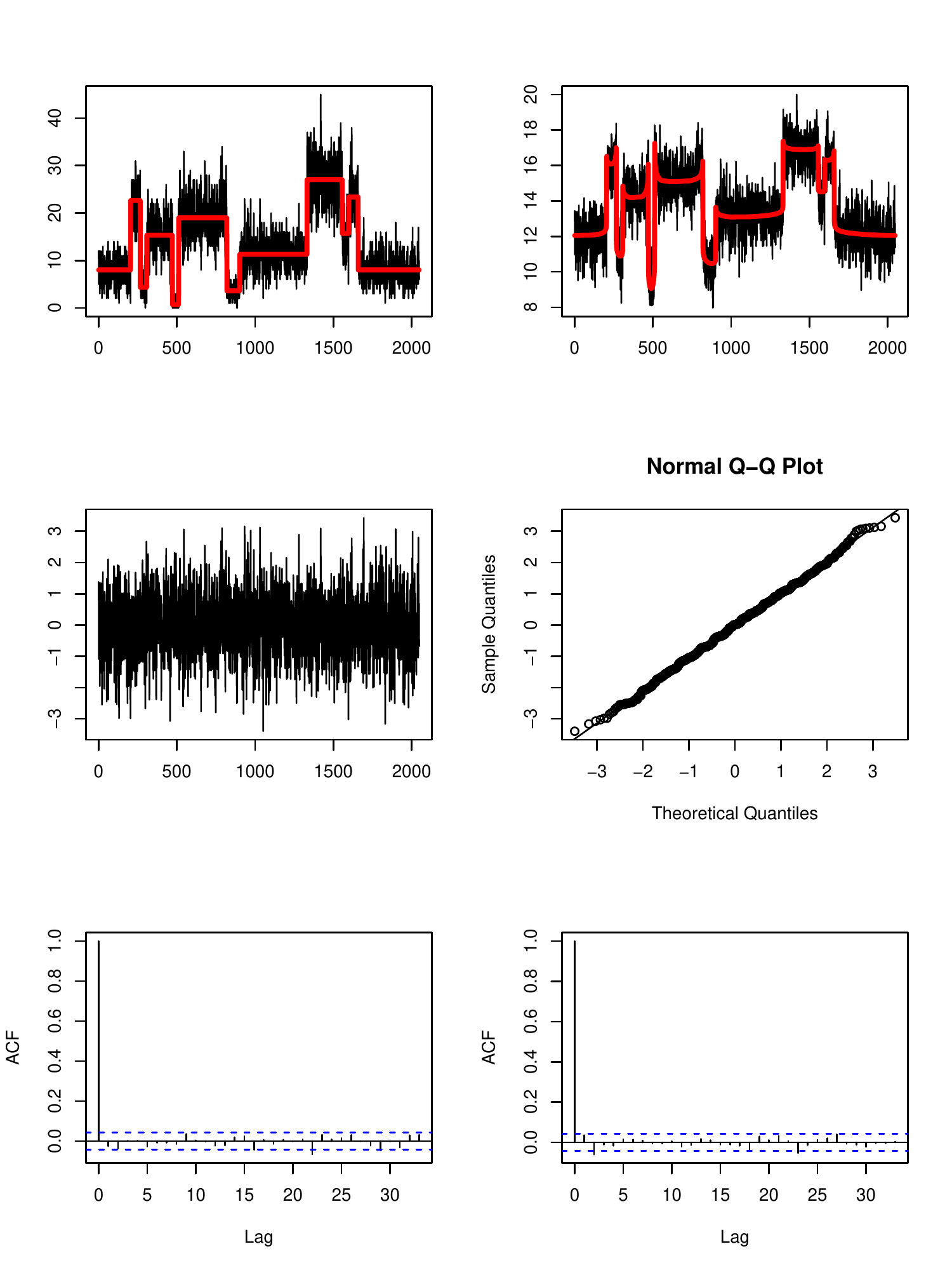}
  \caption{The Poisson model. Top left: Poisson intensity $\Lambda$ (red) and simulated data $\mathbf{X}$ (black).
Top right: the likelihood ratio transform $G(\Lambda)$ (red) and $G(\mathbf{X})$ (black).
Middle left: $G(\mathbf{X}) - G(\Lambda)$. Middle right: Q-Q plot of $G(\mathbf{X}) - G(\Lambda)$ against the normal
quantiles. Bottom left: sample acf plot of $G(\mathbf{X}) - G(\Lambda)$. Bottom right: sample
acf plot of $(G(\mathbf{X}) - G(\Lambda))^2$.}
  \label{fig:gpois}
\end{figure}

\begin{figure}[p]
\centering
  \includegraphics[width=0.95\linewidth]{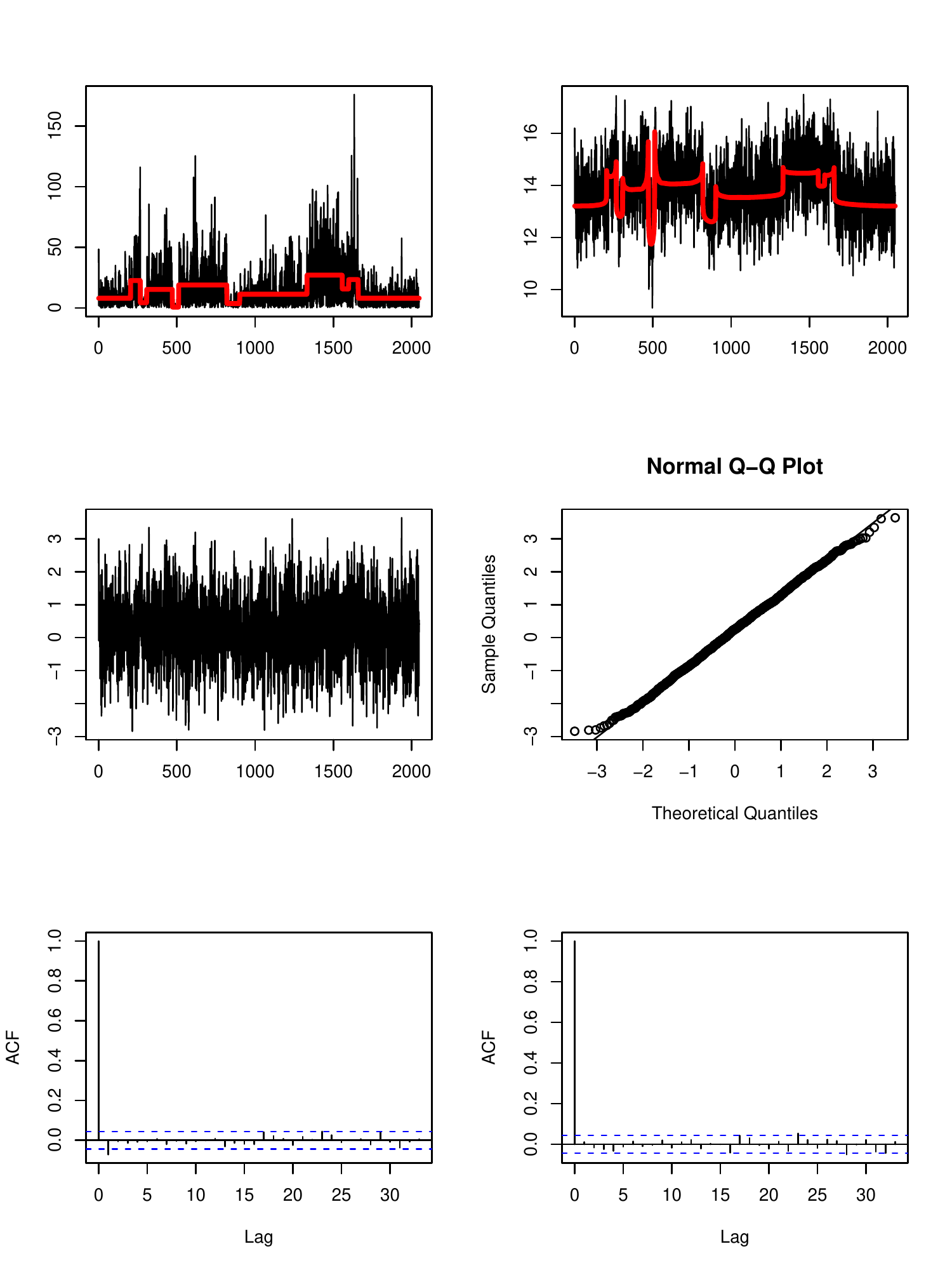}
  \caption{The exponential model. Top left: exponential mean $\sigma^2$ (red) and simulated data $\mathbf{X}$ (black).
Top right: the likelihood ratio transform $G(\sigma^2)$ (red) and $G(\mathbf{X})$ (black).
Middle left: $G(\mathbf{X}) - G(\sigma^2)$. Middle right: Q-Q plot of $G(\mathbf{X}) - G(\sigma^2)$ against the normal
quantiles. Bottom left: sample acf plot of $G(\mathbf{X}) - G(\sigma^2)$. Bottom right: sample
acf plot of $(G(\mathbf{X}) - G(\sigma^2))^2$.}
  \label{fig:gchisq}
\end{figure}

Figures \ref{fig:gpois} and \ref{fig:gchisq} illustrate the results. In both the Poisson and the exponential examples,
the likelihood ratio Haar transform is a very good normalizer and variance-stabilizer: the transformed data minus the
transformed signal shows good agreement with an i.i.d. normal sample; its sample variance equals 1.07 for the Poisson
model and 1.14 for the exponential model. Particularly for the exponential model, the likelihood ratio Haar transform
is a significantly better normalizer than the Haar-Fisz transform (not shown here).

\subsection{California earthquake data}
\label{sec:xquake}

In this section, we revisit the Northern California earthquake dataset, analysed in \cite{fiszhaar} and available from
\verb+http://quake.geo.berkeley.edu/ncedc/catalog-search.html+.
We analyze the time series $N_k$, $k = 1, \ldots, 1024$, where $N_k$ is the number of earthquakes of 
magnitude 3.0 or more which occurred in the $k$th week,
the first week under consideration starting April 22nd, 1981 and the final ending December 5th, 2000.
We assume $N_k \sim \mbox{Pois}(\lambda_k)$ and estimate $\Lambda$ using our likelihood ratio Haar smoother,
used as described in Section \ref{ssec:sm}.

\begin{figure}[t]
\centering
  \includegraphics[width=0.6\linewidth]{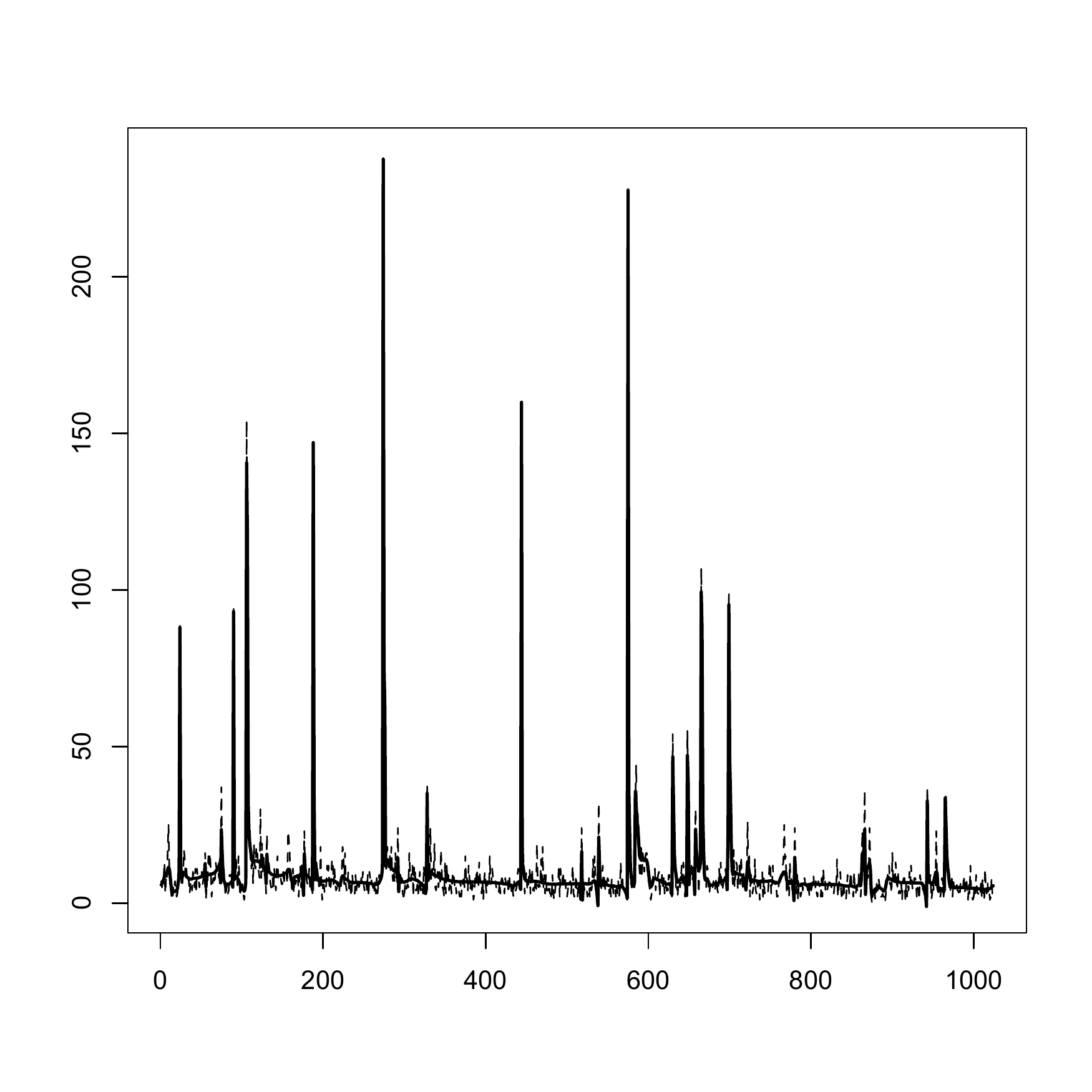}
  \caption{Northern California earthquake data: $N_k$ (dashed) and the likelihood ratio Haar estimate (thick solid). See
  Section \ref{sec:xquake} for details.}
  \label{fig:xquake}
\end{figure}

The estimate and the data are shown in Figure \ref{fig:xquake}. The appearance of the estimator reveals an interesting
phenomenon, not necessarily easily seen in the noisy data: for many of the intensity spikes observed in this dataset,
the intensity in the time period just before the spike appears to be much lower than the intensity in the period following the
spike, which may point to a degree of persistence in the seismic activity following the major spikes in activity
observed in these data.

\section{Additional discussion}
\label{sec:ad}

For unknown distributions, the idea of using empirical likelihood ratio tests (to construct empirical likelihood ratio Haar coefficients)
may seem tempting at first glance, but it is well known that their computation
can be tedious, and therefore they appear to be of little practical use in our context. A more attractive option may be to attempt to estimate, nonparametrically,
the function $f''$ (in the notation of Lemmas \ref{lem:a1}--\ref{lem:mvt}), noting that it is the reciprocal of the function linking the variance to the mean of the
distribution in question. This estimation could be carried out e.g. as described in \cite{f08}. From this, one could attempt
to construct an estimate of the function $f$ itself, which could be used in the computation of the coefficients $g_{j,k}$.

To generalize our likelihood ratio Haar methodology to wavelets other than Haar, one could possibly resort to lifting schemes \citep{s96}, in which the `predict' step could be modified
from simple linear prediction to likelihood-based one.

\appendix

\section{Technical results}

\begin{lemma}
\label{lem:a1}
Let function $f : [u,v] \to \mathbb{R}$ be such that $f'$ is continuous on $[u,v]$ and
$f^{''}$ is continuous on $(u,v)$. There exists a point $\xi \in (u,v)$ such that
\[
f(u) - 2f\left(\frac{u+v}{2}\right) + f(v) = \frac{(u-v)^2}{4} f^{''}(\xi).
\]
\end{lemma}

{\bf Proof.} Let $z = (u+v)/2$ and $\delta = (v-u)/2$, then
\[
f(u) - 2f\left(\frac{u+v}{2}\right) + f(v) = f(z-\delta) - 2f(z) + f(z+\delta).
\]
Defining $g(x) = f(z-x) - 2f(z) + f(z+x)$, Taylor's theorem yields
\begin{equation}
\label{eq:mvt}
g(\delta) = g(0) + \delta g'(0) + \frac{\delta^2}{2} g{''}(\xi') = \frac{\delta^2}{2} g{''}(\xi') =
\frac{\delta^2}{2} \{f{''}(z + \xi') + f{''}(z - \xi')\},
\end{equation}
where $\xi' \in (0, \delta)$. By the intermediate value theorem,
there exists a $\xi \in (z - \xi', z+\xi') \subset [u,v]$ such that
$\{f{''}(z + \xi') + f{''}(z - \xi')\}/2 = f{''}(\xi)$, which by
(\ref{eq:mvt}) completes the result.

\vspace{10pt}

\begin{lemma}
\label{lem:grt}
Let function $f : [u,v] \to \mathbb{R}$ be such that $f'$ is continuous on $[u,v]$ and
$f^{''}$ is convex on $(u,v)$. Then
\[
f(u) - 2f\left(\frac{u+v}{2}\right) + f(v) \ge \frac{(u-v)^2}{4} f^{''}\left( \frac{u+v}{2} \right).
\]
\end{lemma}

{\bf Proof.} Straightforward from the convexity of $f{''}$ and (\ref{eq:mvt}).

\vspace{10pt}

\begin{lemma}
Let function $f : [u,v] \to \mathbb{R}$ be such that $f'$ is continuous on $[u,v]$ and
$f^{''}$ is nonincreasing on $[u,v)$. Then
\[
f(u) - 2f\left(\frac{u+v}{2}\right) + f(v) \le \frac{(u-v)^2}{8} \left\{ f{''}\left( \frac{u+v}{2} \right) + f{''}(u) \right\}.
\]
\end{lemma}

{\bf Proof.} Straightforward from (\ref{eq:mvt}) and the fact that $f{''}$ is nonincreasing on $[u,v)$.

\vspace{10pt}

\begin{lemma}
\label{lem:mvt}
Let function $f : [u,v] \to \mathbb{R}$ be such that $f'$ is continuous on $[u,v]$ and
$f^{''}$ is convex on $[u,v]$. Then
\[
f(u) - 2f\left(\frac{u+v}{2}\right) + f(v) \le \frac{(u-v)^2}{8} \left\{ f{''}(v) + f{''}(u) \right\}.
\]
\end{lemma}

{\bf Proof.} Straightforward from the convexity of $f{''}$ and (\ref{eq:mvt}).

\vspace{10pt}

\begin{lemma}
\label{lem:cramer}
The Poisson distribution satisfies Cramer's conditions.
\end{lemma}

{\bf Proof.} The Poisson distribution is log-concave, and \cite{ss11}, Lemma 7.4,
show that all log-concave random variables $Z$ are central moment bounded with real parameter $L>0$, that is, satisfy for 
any integer $i \ge 1$,
\[
E|Z - E(Z)|^i \le i\,L\,E|Z - E(Z)|^{i-1}.
\]
Moreover, again by \cite{ss11}, Lemma 7.5, we have
\[
L = 1 + \max(E(|Z - E(Z)|\quad |\quad Z \ge E(Z)),\quad E(|Z - E(Z)|\quad |\quad Z < E(Z))),
\]
which for the Pois($\lambda$) distribution gives $L = O(\lambda^{1/2})$. But
\begin{eqnarray*}
E|Z - E(Z)|^i & \le & i\,L\,E|Z - E(Z)|^{i-1}\\
& \le & i! L^{i-2} E(Z - E(Z))^2,
\end{eqnarray*}
which completes the proof of the lemma.

\vspace{10pt}

{\bf Proof of Theorem \ref{th:pois}.}

We first show that $P({\mathcal A} \cap {\mathcal B}) \to 1$. We have
\begin{equation}
\label{eq:bonfineq}
P({\mathcal A}^c) \le \sum_{j=J_0+1}^J \sum_{k=1}^{2^{J-j}} P((\bar{\lambda}_{(k-1)2^j+1}^{k2^j})^{-1/2}|d_{j,k} - \mu_{j,k}| \ge t_1).
\end{equation}
Since by Lemma \ref{lem:cramer}, the Poisson distribution satisfies Cramer's conditions, $\Lambda$ is bounded from above and away from zero,
and $2^{J_0} = O(n^{\beta})$ for $\beta \in (0,1)$, the strong asymptotic normality from the Corollary underneath the proof of Theorem 1 in
\cite{rss78}
can be used, which in our context implies that if $t_1 = O(\log^{1/2}n)$, then
\begin{equation}
\label{eq:lith}
P((\bar{\lambda}_{(k-1)2^j+1}^{k2^j})^{-1/2}|d_{j,k} - \mu_{j,k}| \ge t_1) \le C \Phi(t_1),
\end{equation}
where $\Phi(\cdot)$ is the cdf of the standard normal distribution and $C$ is a universal constant. Using (\ref{eq:lith}), Mills' ratio inequality and the fact that
$t_1 = C_1 \log^{1/2} n$, we bound (\ref{eq:bonfineq}) from above by $\tilde{C} \log^{-1/2} n\,\, n^{1-\beta - C_1^2 / 2}$, where $\tilde{C}$ is a constant,
which proves that $P({\mathcal A}) \to 1$. The proof that $P({\mathcal B}) \to 1$ is identical.

We now turn to the estimator. Due to the orthonormality of the Haar transform, we have
\begin{equation}
\label{eq:se}
n^{-1}\| \hat{\Lambda} - \Lambda  \|^2 = n^{-1} \sum_{j=1}^J \sum_{k=1}^{2^{J-j}} (\hat{\mu}_{j,k} - \mu_{j,k})^2 +
n^{-1} (s_{J,1} - \tilde{\lambda})^2,
\end{equation}
where $\tilde{\lambda} = n^{-1/2} \sum_{k=1}^n \lambda_k$.

We first consider scales $j = 1, \ldots, J_0$, for which $\hat{\mu}_{j,k} = 0$. At each scale $j$,
there are at most $N$ indices $k$ for which $\mu_{j,k} \neq 0$.
From the definition of $d_{j,k}$, for those $\mu_{j,k}$, we have $\mu_{j,k} \le 2^{j/2-1} \Lambda'$, which gives
\begin{equation}
\label{eq:zero}
\sum_{j=1}^{J_0} \sum_{k=1}^{2^{J-j}} (\hat{\mu}_{j,k} - \mu_{j,k})^2 \le N (\Lambda')^2 \sum_{j=1}^{J_0} 2^{j-2} = N (\Lambda')^2 (2^{J_0-1} - \frac{1}{2}).
\end{equation}

We now consider the remaining scales $j = J_0+1, \ldots, J$ and first take an arbitrary index $(j,k)$ for which $\lambda_i$ is
not constant for $i = (k-1)2^j+1, \ldots, k2^j$. For such a $(j,k)$, we have (using Lemma \ref{lem:grt} in the second inequality)
\begin{eqnarray*}
(\hat{\mu}_{j,k} - \mu_{j,k})^2 & = & (d_{j,k}\mathbb{I}(|g_{j,k}| > t) - \mu_{j,k})^2\\
& \le & 2d_{j,k}^2\mathbb{I}(|g_{j,k}| \le t) + 2(d_{j,k} - \mu_{j,k})^2\\
& \le & 2d_{j,k}^2\mathbb{I}(|d_{j,k}| \le t (\bar{X}_{(k-1)2^j+1}^{k2^j})^{1/2}   ) + 2(d_{j,k} - \mu_{j,k})^2\\
& \le & 2 t^2 \bar{X}_{(k-1)2^j+1}^{k2^j} + 2(d_{j,k} - \mu_{j,k})^2\\
& \le & 2 t^2 (\bar{\lambda}_{(k-1)2^j+1}^{k2^j} + t_2 2^{-j/2}(\bar{\lambda}_{(k-1)2^j+1}^{k2^j})^{1/2}) + 2 t_1^2 \bar{\lambda}_{(k-1)2^j+1}^{k2^j}.
\end{eqnarray*}
Summing the bound over the at most $N$ indices $k$ within each scale for which
$\lambda_i$ is not constant for $i = (k-1)2^j+1, \ldots, k2^j$, as well as over scales
$j = J_0+1, \ldots, J$, and noting that $\bar{\lambda}_{(k-1)2^j+1}^{k2^j} \le \bar{\Lambda}$, gives the upper bound of
\begin{equation}
\label{eq:jumps}
2N\bar{\Lambda}^{1/2} \left\{   (J-J_0)(t^2 + t_1^2) \bar{\Lambda}^{1/2} + t^2 t_2 (1 + 2^{-1/2}) 2^\frac{-J_0+1}{2} \right\}.
\end{equation}

We finally consider again the scales $j = J_0+1, \ldots, J$ and those indices $(j,k)$ for which 
$\lambda_i$ is constant for $i = (k-1)2^j+1, \ldots, k2^j$, which implies $\mu_{j,k} = 0$. For each such $(j,k)$,
we have
\begin{eqnarray*}
(\hat{\mu}_{j,k})^2 & = & d_{j,k}^2\mathbb{I}(|g_{j,k}| > t).
\end{eqnarray*}
Consider the following sequence of inequalities, with the first one being implied by Lemma \ref{lem:mvt}, and the second using the fact
that $\bar{\lambda}_{(k-1)2^j+1}^{(k-1)2^j+2^{j-1}} = \bar{\lambda}_{(k-1)2^j+2^{j-1}+1}^{k2^j} = \bar{\lambda}_{(k-1)2^j+1}^{k2^j}$.
\begin{eqnarray}
|g_{j,k}| > t & \Rightarrow & |d_{j,k}| 2^{-1/2} \left|  \frac{1}{\bar{X}_{(k-1)2^j+1}^{(k-1)2^j+2^{j-1}}} + \frac{1}{\bar{X}_{(k-1)2^j+2^{j-1}+1}^{k2^j}}  \right|^{1/2} > t\nonumber\\
& \Rightarrow & \frac{|d_{j,k}|}{(\bar{\lambda}_{(k-1)2^j+1}^{k2^j} - \delta)^{1/2}} > t\quad \lor\quad |\bar{X}_{(k-1)2^j+1}^{(k-1)2^j+2^{j-1}} - \bar{\lambda}_{(k-1)2^j+1}^{(k-1)2^j+2^{j-1}}| \ge \delta
\nonumber\\
&& \lor\quad |\bar{X}_{(k-1)2^j+2^{j-1}+1}^{k2^j} - \bar{\lambda}_{(k-1)2^j+2^{j-1}+1}^{k2^j}| \ge \delta\nonumber\\
& \Leftrightarrow & \frac{|d_{j,k}|}{(\bar{\lambda}_{(k-1)2^j+1}^{k2^j})^{1/2}} > t
\left( 1 - \frac{\delta}{\bar{\lambda}_{(k-1)2^j+1}^{k2^j}}  \right)^{1/2}\nonumber\\
&& \lor\quad 2^{j/2} (\bar{\lambda}_{(k-1)2^j+1}^{(k-1)2^j+2^{j-1}})^{-1/2} |\bar{X}_{(k-1)2^j+1}^{(k-1)2^j+2^{j-1}} - \bar{\lambda}_{(k-1)2^j+1}^{(k-1)2^j+2^{j-1}}| \ge \delta
 2^{j/2} (\bar{\lambda}_{(k-1)2^j+1}^{(k-1)2^j+2^{j-1}})^{-1/2}\nonumber\\
&& \lor\quad 2^{j/2} (\bar{\lambda}_{(k-1)2^j+2^{j-1}+1}^{k2^j})^{-1/2} |\bar{X}_{(k-1)2^j+2^{j-1}+1}^{k2^j} - \bar{\lambda}_{(k-1)2^j+2^{j-1}+1}^{k2^j}| \ge\nonumber\\
\label{eq:impl}
&&\quad\quad \delta 2^{j/2} (\bar{\lambda}_{(k-1)2^j+2^{j-1}+1}^{k2^j})^{-1/2}.
\end{eqnarray}
Let us set $\delta = t_2 2^{-j/2} (\bar{\lambda}_{(k-1)2^j+1}^{k2^j})^{1/2}$, then if
\begin{equation}
\label{eq:condt12}
t_1 \le t (1 - t_2 2^{-j/2} (\bar{\lambda}_{(k-1)2^j+1}^{k2^j})^{-1/2})^{1/2},
\end{equation}
then the right-hand side of the implication (\ref{eq:impl}) is negated on ${\mathcal A} \cap {\mathcal B}$, which implies that so is the left-hand side, and therefore 
$\hat{\mu}_{j,k} = 0$. Note (\ref{eq:condt12}) is satisfied if (\ref{eq:condt21}) holds.

Putting together (\ref{eq:zero}) and (\ref{eq:jumps}) and noting that $n^{-1} (s_{J,1} - \tilde{\lambda})^2 \le n^{-1} t_1^2 \bar{\lambda}_1^n$
on ${\mathcal A}$, we bound (\ref{eq:se}) by
\[
\frac{1}{2}n^{-1} N (\Lambda')^2 (n^\beta - 1) +
2n^{-1}N\bar{\Lambda}^{1/2} \left\{   (J-J_0)(t^2 + t_1^2) \bar{\Lambda}^{1/2} + t^2 t_2 (2 + 2^{1/2})
n^{-\beta/2} \right\} + n^{-1} t_1^2 \bar{\lambda}_1^n
\]
on condition that (\ref{eq:condt21}) holds, which completes the proof.

\section*{Acknowledgements}

Piotr Fryzlewicz's work was supported by the Engineering and
Physical Sciences Research Council grant no. EP/L014246/1.
Data products for this study were accessed through the Northern California Earthquake Data Center (NCEDC), doi:10.7932/NCEDC.

\bibliographystyle{plainnat}
\bibliography{pio}

\end{document}